\documentclass[namedreferences,a4paper]{solarphysics}

\usepackage[hyperref,optionalrh]{spr-sola-addons} 
\usepackage{epsfig}          
\usepackage{graphicx}        
\usepackage{courier}         
\usepackage{color}           
\usepackage{amsmath}
\usepackage{enumitem}   

\chardef\us=`\_

\begin{document}

\begin{opening}

\title{Working Principle of the Calibration Algorithm for High Dynamic Range Solar Imaging with the \textit {Square Kilometre Array} Precursor}

\author[addressref={National Centre for Radio Astrophysics, Tata Institute of Fundamental Research, S.P.Pune University Campus, Pune 411007,India},email={dkansabanik@ncra.tifr.res.in, devojyoti96@gmail.com}]{\inits{D.}\fnm{Devojyoti}~\lnm{Kansabanik}\orcid{0000-0001-8801-9635}}

\runningauthor{D. Kansabanik}
\runningtitle{Working Principle of AIRCARS}

\begin{abstract}
Imaging the low-frequency radio Sun is an intrinsically challenging problem. Meter-wavelength solar emission spans angular scales from a few arcminutes to a few degrees. These emissions show temporal and spectral variability on a sub-second and sub-MHz scales. The brightness temperature of these emissions also varies by many orders of magnitude, which requires high-dynamic-range spectroscopic snapshot imaging. With the unique array configuration of the \textit {Murchison Widefield Array} (MWA), and the robust calibration and imaging pipeline, \textit{Automated Imaging Routine for the Compact Arrays for the Radio Sun} (AIRCARS) produces the best spectroscopic snapshot solar images available to date. The working principle and the strength of this algorithm are demonstrated using statistical analysis and simulation. AIRCARS uses the partial phase stability of the MWA, which has a compact core with a large number of antenna elements distributed over a small array footprint. The strength of this algorithm makes it a state-of-the-art calibration and imaging pipeline for low-frequency solar imaging, which is expected to be highly suitable for the upcoming \textit {Square Kilometre Array} (SKA) and other future radio interferometers for producing high-dynamic-range and high-fidelity images of the Sun.     
\end{abstract}
\end{opening}

\section{Introduction}
\begin{figure}
    \centering
    \includegraphics[trim={0cm 2.5cm 0.5cm 0cm},clip,scale=0.24]{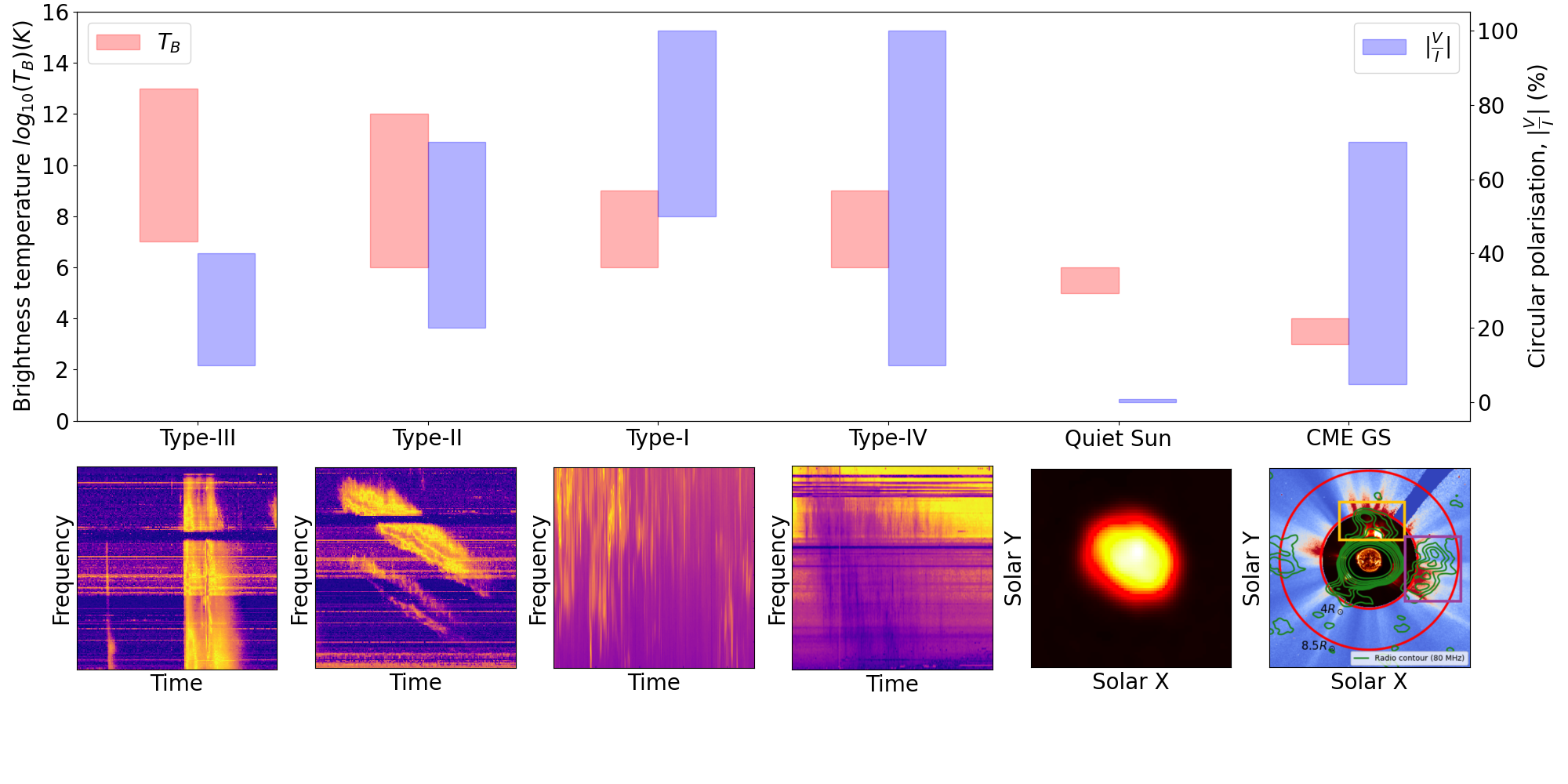}
    \caption{ \textit{Top panel :} The expected range of brightness temperature [$T_\mathrm{B}$] and circular polarization fraction for different kinds of low-frequency solar radio emissions are shown by the {\it blue} and {\it red} bars respectively. \textit{Bottom panel :} Sample dynamic spectra for Type-I, -II, -III, and -IV radio bursts. Type-II, -III, and -IV dynamic spectra have been obtained from the \textit {Learmonth Radio Spectrograph}. The Type-I dynamic spectrum is obtained from the \textit{Murchison Widefield Array} (MWA). Dynamic spectra have different spectro-temporal structures spanning a large range of spectral and temporal widths. The images of the \textit {last two panels} show the quiet-Sun emission and gyrosynchrotron emission from a CME. These emissions have spatial structures spanning a large range of angular scales. The images of the quiet-Sun and CME are from MWA.}
    \label{fig:different_emissions}
\end{figure}

The solar emission covers the entire electromagnetic spectrum starting from $\gamma$-rays to radio wavelengths. Different physical mechanisms produce emissions at different wavelengths. Several emission mechanisms such as plasma emission, thermal bremsstrahlung, and gyrosynchrotron produce the meter-wavelength solar emission at the solar corona. Low-frequency radio observations are particularly important to measure the coronal magnetic fields and the non-thermal electron population, which are rather hard to do using observations at other wavelengths. Despite its well-appreciated importance, low-frequency imaging observation of the Sun is one of the least explored areas of solar physics. The brightness temperature [$T_\mathrm{B}$] of the low-frequency solar emissions can vary from $\approx10^3-10^4\ \mathrm{K}$ for gyrosynchrotron emission from CME plasma \citep{bastian2001,Mondal2020a} to $\approx10^{13}\ \mathrm{K}$ for bright Type-III radio bursts \citep{McLeanBook,Reid2014} (shown by red bars in Figure \ref{fig:different_emissions}) over a background quiescent $T_\mathrm{B}$ of $\approx10^6\ \mathrm{K}$. Depending upon the emission mechanism at play, the polarization fraction of the meter wavelength solar emission can also vary from $\leq 1\ \%$ to $\approx100\ \%$ \citep{McLeanBook,Nindos2020} (shown by the blue bars in Figure \ref{fig:different_emissions}). 

Imaging the Sun at low radio frequencies with high fidelity is an intrinsically challenging problem. The Sun is an extended source having morphology spanning a large range of angular scales, from a few degrees to a few arcminutes at meter wavelengths. The meter-wavelength solar emission varies over small temporal and spectral scales, which imposes a requirement for snapshot spectroscopic imaging. The need to be able to see features varying vastly in $T_\mathrm{B}$ highlights the need for a high imaging dynamic range. It has only recently become possible to meet these exacting requirements for solar radio imaging using the \textit{Murchison Widefield Array} \citep[MWA:][]{Tingay2013, Wayth2018} operating at $80 \,--\, 300 \ \mathrm{MHz}$ with an instantaneous bandwidth of $30.72\ \mathrm{MHz}$. The MWA has a large number of antenna elements distributed over a small array footprint and is especially well suited for snapshot spectroscopic imaging. To produce high-dynamic-range high-fidelity spectroscopic snapshot solar images from the MWA solar observation, \citet{Mondal2019} developed a novel calibration and imaging pipeline called \textit{Automated Imaging Routine for Compact Arrays for the Radio Sun} (AIRCARS). 

AIRCARS has been used on a set of MWA solar observations in different solar conditions, and successfully produced the best spectroscopic snapshot images of the Sun at low frequencies obtained to date. The dynamic range (DR) of the images produced by AIRCARS varies between $>300$ to about $10^5$. It has led to many new discoveries over the last few years \citep{Mohan2019a,Mohan2019b,Mondal2020a,Mondal2020b,Mondal2021a,Mohan2021a,Mohan2021b}. \cite{Mondal2019} described the implementation of AIRCARS and also demonstrated that it can even perform the calibration without any dedicated calibrator observation. However the explanation provided by \citet{Mondal2019} as to why the algorithm works was rather limited. Additionally, due to some unappreciated aspects that have been understood since then, the way the algorithm works is different from the description they provided. Here this difference is discussed and the working principle of AIRCARS is discussed using statistical and quantitative analysis.

\section{Suitability of the Array Configuration of the MWA for High-Fidelity Spectroscopic Snapshot Solar Imaging}\label{sec:suitability_of_MWA}
Radio interferometric imaging is a Fourier imaging technique \citep{McCready1947,thompson2017}. A radio interferometer is made up of multiple antenna elements (or dishes) distributed over the ground. Each antenna element of the array receives radio emission from the sky and converts it into electronic voltage. The cross-correlation of the measured voltages between the antenna pairs is known as {\it visibilities}. Each of these visibilities corresponds to a single Fourier component of the sky-brightness distribution in a two-dimensional Fourier plane, which is known as the {\it uv-}plane. The inverse Fourier transform of the measured visibilities on the {\it uv-}plane gives the true sky-brightness distribution. Ideally, the {\it uv-}plane has to be sampled at the spatial Nyquist resolution. Most of the conventional radio interferometers, like the VLA \citep{VLA2009}, uGMRT \citep{Gupta_2017}, WSRT \citep{WSRT2021}, LOFAR \citep{lofar2013}, etc, have a limited number of antenna elements distributed sparsely over a large area on the ground. Hence, the instantaneous sampling of the {\it uv-}plane is very sparse and does not meet the Nyquist criteria. 

One way to sample the {\it uv-}plane densely is the so-called “large-N” array configuration. The MWA array design follows the ``large-N" array configuration. It has 128 antenna elements distributed over a small array footprint and provides dense spectroscopic snapshot {\it uv-}coverage. The MWA has two phases of operation. The MWA Phase-I \citep{Tingay2013}, has a compact condensed core of $\approx1.5\ \mathrm{km}$ in diameter with a quasi-random distribution of antenna elements. The other antenna elements are distributed over a region up to $3\ \mathrm{km}$ in diameter. The MWA Phase-II \citep{Wayth2018} has two configurations : ``compact" and ``extended". The Phase-II extended configuration has baselines up to $\approx5.3\ \mathrm{km}$ in diameter and the compact configuration has the maximum useful baseline for snapshot imaging up to $\approx0.4\ \mathrm{km}$ in diameter. The radio Sun has an angular diameter of $\approx 40 - 60\ \mathrm{arcmin}$ at $150\, \mathrm{MHz}$. The size of the point spread function (PSF) for the compact configuration $\approx$20 $\mathrm{arcmin}$ at $150\, \mathrm{MHz}$. Hence, barely 8 PSFs can be fitted inside the radio Sun at these frequencies, and the compact configuration is not favorable for solar observation. On the other hand, the Phase-I provides angular resolution of $\approx2.5\ \mathrm{arcmin}$ and the Phase-II extended configuration provides angular resolution of $\approx1.5\ \mathrm{arcmin}$ at 150 $\mathrm{MHz}$, and the most favorable configurations for solar observations. 

\begin{figure*}
    \centering
    \includegraphics[trim={2.3cm 1.5cm 3cm 1.5cm},clip,scale=0.48]{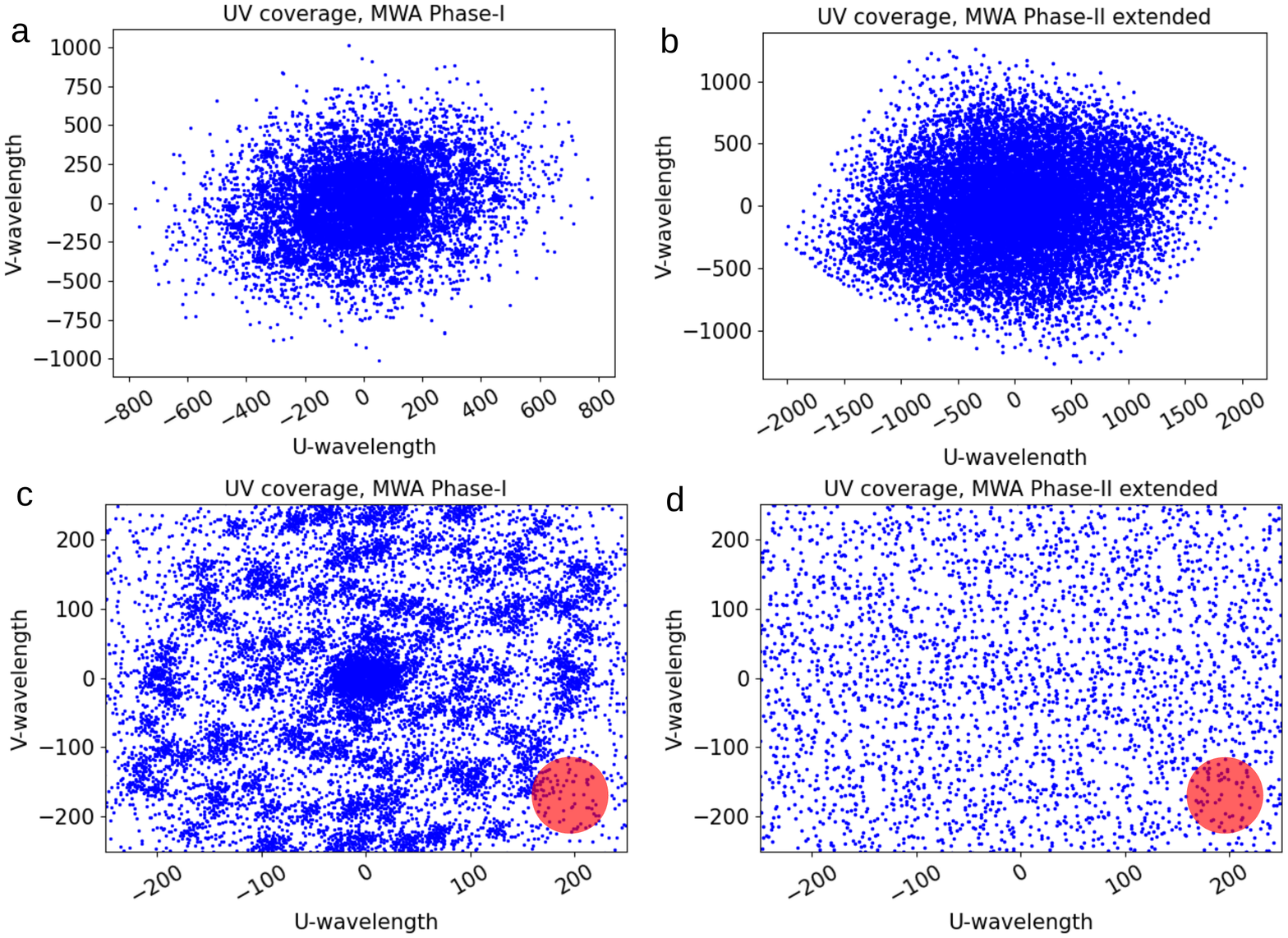}
    \includegraphics[trim={2.5cm 14.5cm 4.5cm 1.8cm},clip,scale=0.37]{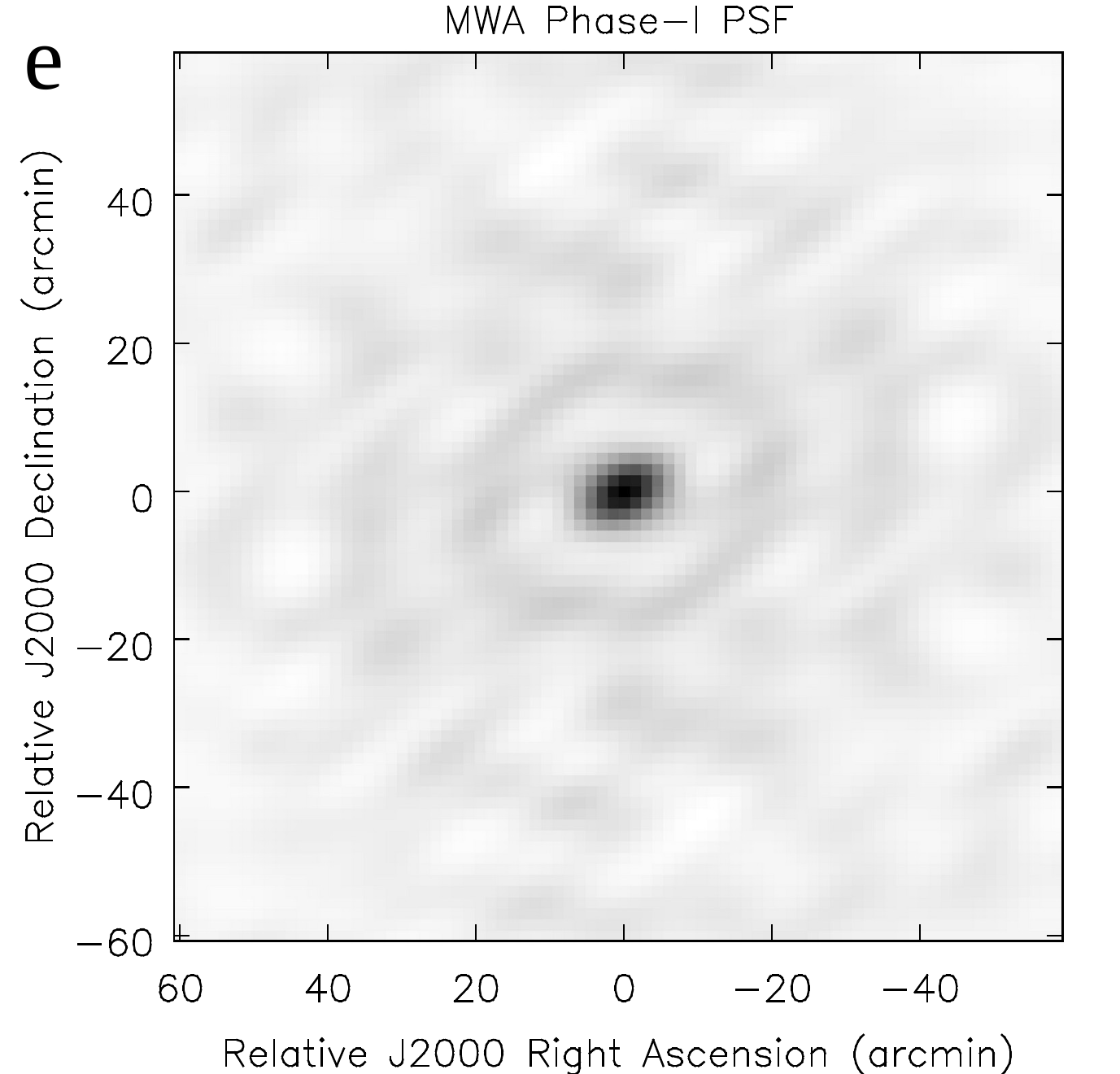}\includegraphics[trim={2.5cm 14.5cm 2.2cm 1.8cm},clip,scale=0.37]{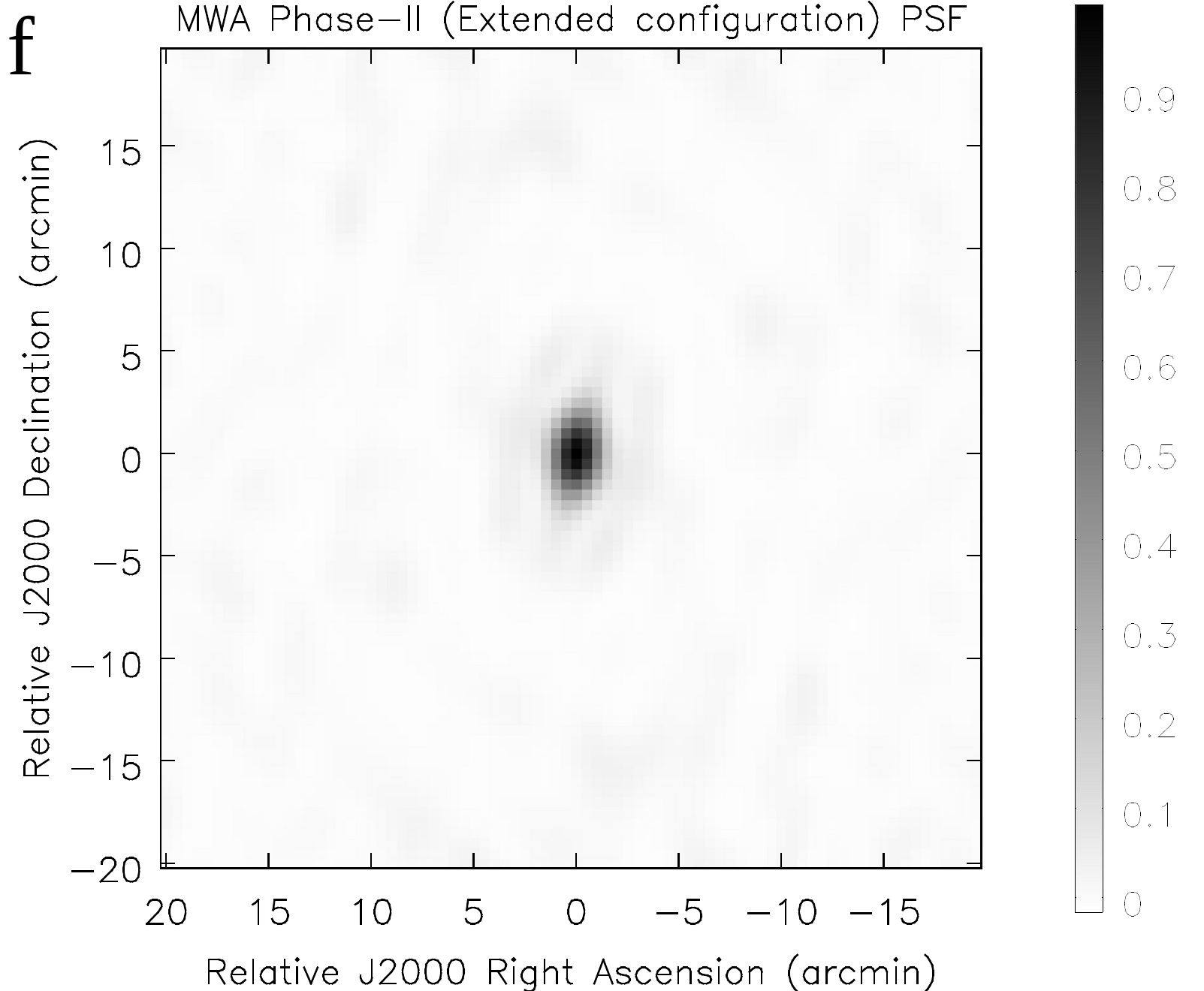}
    \caption{{\bf Snapshot {\it uv-}coverage and point spread function (PSF) of the MWA  at 150 MHz}. (a)  Snapshot {\it uv-}coverage of MWA Phase-I, and {\bf (b)} Phase-II extended configuration. {\bf (c,d)} Zoomed-in version of the {\it uv-}coverage over a region of size $100\ \lambda$. \textit {Red circles} correspond to the {\it uv-}cell for a source with 1$^\circ$ angular scale.  {\bf (e)} Un-tapered and naturally weighted PSF of MWA Phase-I, and {\bf (f)} Phase-II extended configuration at 150 MHz.}
    \label{fig:uvcoverage}
\end{figure*}

The snapshot {\it uv-}coverage of MWA Phase-I and Phase-II extended configurations are shown in Figure \ref{fig:uvcoverage}a and b respectively. The zoomed-in versions over a square of $250 \lambda$ are shown in Figure \ref{fig:uvcoverage}c and d respectively. The red circle shows the {\it uv-}cell required for Nyquist sampling for a source with $1^\circ$ angular scale, which is the approximate angular size of the Sun at the meter-wavelengths. It is evident from Figure \ref{fig:uvcoverage}c and d that, the density of {\it uv-}sampling approaches or even exceeds the Nyquist criterion over a significant part of the uv-plane. The bottom-left panel of Figure \ref{fig:uvcoverage} shows the naturally weighted and un-tapered spectroscopic snapshot PSF for the Phase-I and the right panel is for the Phase-II extended configuration. These snapshot PSFs are extremely well-structured, which reduces the deconvolution artifacts in the final images. These properties of the MWA array configuration make it well-suited for the high-dynamic-range spectroscopic snapshot solar imaging. 

\section{A Brief Overview of Radio Interferometric Calibration}\label{sec:calibration}
The true source visibility [$V_{\mathrm{pq}}$] between a antenna pairs, $p$ and $q$, is corrupted by the complex instrumental gains and due to the atmospheric propagation effects. At low radio frequencies, the ionospheric propagation effect is the major atmospheric propagation effect. In practice, both the instrumental gain and ionospheric propagation effect are merged into a single complex gain term. The measured visibility [$V_{\mathrm{pq}^\prime}$] can be written in terms of the $V_\mathrm{{pq}}$ as
\begin{equation}\label{eq:measurement_eq}
\begin{split}
     V_{\mathrm{pq}}^\prime(\nu,\ t,\ {\bf l}) &= J_\mathrm{p}(\nu,\ t,\ {\bf l})\ V_{\mathrm{pq}(\nu,\ t,}\ {\bf l})\ J_\mathrm{q}^\dagger(\nu,\ t,\ {\bf l}) + N_\mathrm{{pq}}\\
      &= |J_\mathrm{p}(\nu,\ t,\ {\bf l})|\ V_{\mathrm{pq}}(\nu,\ t,\ {\bf l})\ |J_\mathrm{q}^\dagger(\nu,\ t,\ {\bf l})| \mathrm{e}^{i\left[\phi_\mathrm{p}(t)-\phi_\mathrm{q}(t)\right]} \\&+ N_\mathrm{pq}
\end{split}
\end{equation}
where, $J_\mathrm{p}(\nu,\ t,\ {\bf l})$ and $J_\mathrm{q}(\nu,\ t,\ {\bf l})$ are the complex gain terms incorporating both the instrumental and the ionospheric effects, $N_{\mathrm{pq}}$ is the additive noise, $\nu,\ t,\ {\bf l}$ represent the observing frequency, time and direction in the sky plane respectively, and $\phi_\mathrm{p}$ and $\phi_\mathrm{q}$ are the phase parts of $J_\mathrm{p}$ and $J_\mathrm{q}$ respectively. Equation \ref{eq:measurement_eq} is popularly known in the literature as the {\it measurement equation} \citep{Hamaker1996_1} for a radio interferometer.  One has to estimate $J_\mathrm{p}(t,\ \nu,\ {\bf l})=G_\mathrm{p}(t)B_\mathrm{p}(\nu)E_\mathrm{p}({\bf l})$ for all the antenna elements and correct them to obtain $V_\mathrm{{pq}}$ from the $V_{\mathrm{pq}}^\prime$. $J_\mathrm{p}(\nu,\ t,\ {\bf l})$ can be decomposed into two major parts :

\begin{enumerate}[label=\roman*)]
\item{\bf Direction-independent terms -} $G_\mathrm{p}(t)$ and $B_\mathrm{p}(\nu)$ are the two direction-independent components of $J_\mathrm{p}$. $G_\mathrm{p}(t)$ represents the time-variable instrumental and ionospheric gain and $B_\mathrm{p}(\nu)$ is the instrumental bandpass.

\item{\bf Direction-dependent terms -} Direction-dependent effects arise for an array with a large field of view (FoV) \citep{Lonsdale2005}. Propagation of radio emission from different parts of the sky through different parts of the ionosphere introduces direction dependent complex gain [$E_\mathrm{p}({\bf l})$].
\end{enumerate}

The standard practice in radio interferometric calibration is to observe a calibrator source with known flux density, spatial structure, and spectral properties, and use it to estimate the $G_\mathrm{p}(t)$ and $B_\mathrm{p}(\nu)$. For wide FoV instruments, instead of a single calibrator source, a global sky model is also used to estimate the direction-dependent gain term [$E_\mathrm{p}({\bf l})$].

\section{Challenges of Solar Observation With the MWA}\label{sec:challenges}
Being an aperture array instrument, the MWA poses several challenges for solar observation. Hence, the standard calibration methods are not applicable for solar observations with the MWA. The challenges and differences are as follows :

\begin{enumerate}[label=\roman*)]
    \item The MWA has a large FoV. Based on the FWHM of the primary beam, at 150 {$\mathrm{MHz}$} the FoV of the MWA is $\approx$610 $\ \mathrm{degree^2}$ \citep{Tingay2013}. Being an aperture array instrument, the primary beam sidelobes of the MWA are very high; $\approx\ 10\%$ \citep{Sokowlski2017, Line2018}. Hence, any calibrator observations during the daytime are corrupted by the solar flux.
    \item At the MWA, calibrators are routinely observed either before sunrise or after sunset and used to determine antenna gains of the array.  
    \item During daytime, the ambient temperature of the surroundings increases, and increases the receiver temperature. This increase in temperature changes the length of the cables connecting the antennas and introduces a different amount of additional phase to the different antennas.
    \item The daytime ionosphere \citep{mondal_ionosphere} and night time \citep[e.g.][etc.]{Loi2015a,Loi2015b,Walker2018} over the MWA array can be significantly different.
    \item Both temperature increase and different ionospheric phases cause a significant difference between the complex gains during daytime and that obtained at nighttime. These differences in the phases of the antenna gains are shown in Figure \ref{fig:cal_aircars_gain_diff}. The difference can be as large as $\approx80\ \mathrm{degrees}$. 
    \item Due to this significant difference between antenna phases, nighttime calibration solutions do not necessarily provide a good starting point for the daytime solar observations.
\end{enumerate}
Moreover, during the initial phase of its operation (2013 to 2015), most of the solar observations with the MWA did not have a dedicated nighttime calibrator observation with the same spectral configuration as the solar observation. These require relying on a self-calibration-based approach for MWA solar observation.

\subsection{Requirement of Direction-Dependent Calibration}\label{sec:direction_dependent_calibration}
For wide FoV instruments like the MWA, direction-dependent calibration is necessary, and it is implemented in the standard calibration and image-processing pipeline for the MWA : Real Time System \citep[RTS:][]{Mitchell2008}. But, the Sun is the source with the highest flux density in the low-frequency radio sky. The flux density of even the quiet Sun is more than $10^4\ \mathrm{Jy}$, which can increase by a few orders of magnitudes during active emission. On the other hand, the flux densities of only a handful sources lie in the range of hundreds of Jy and that for the vast majority of sources lie in the range of a Jy and lower. This effectively reduces the solar observation to a small FoV problem, with a single bright source at the phase center that dominates the overall visibility. Hence, for MWA solar observation, direction-dependent calibration is not required and is not implemented in AIRCARS.

\section{Tackling the Challenges : A Brief Description of AIRCARS}\label{sec:aircars_description}
AIRCARS is an automated and robust self-calibration-based calibration and imaging pipeline for the MWA solar observation \citep[][M19 hereafter]{Mondal2019}. A brief overview of the AIRCARS is described below.
\subsection{M19 Algorithm}\label{subsec:M19}
The M19 algorithm is as follows : 
\begin{enumerate}[label=\roman*)]
    \item When a dedicated nighttime calibrator observation is available for solar observation, calibration solutions obtained from nighttime calibrators are applied first. If this is not available, AIRCARS starts the calibration using the source model made from the uncalibrated observed visibilities.
    \item Choose the visibilities only between core antennas (shown by the blue points in Figure 2 of M19) of the MWA and make a low-resolution source model of the Sun.
    \item Perform phase-only gain calibration using the source model, apply the gain solutions, and make an improved source model.
    \item When the DR has converged, antennas with increasing distance from the core are added in small steps. These additional antennas do not have the gain solutions from the previous self-calibration rounds. 
    \item One round of phase-only self-calibration is performed after the addition of the new antennas.
    \item When all antennas are added in the self-calibration, AIRCARS starts amplitude and phase self-calibration with all antennas taken into consideration. This process continues until the DR of the image has converged.
    \item AIRCARS uses a certain convergence in the dynamic range given by the user. There is a minimum number (about five) of fixed iterations after the amplitude and phase self-calibrations, after which the convergence is checked. This has been done to avoid some local convergence.
\end{enumerate}

\subsection{Implemented Algorithm}\label{subsec:aircars_new}
AIRCARS uses the Common Astronomy Software Application \citep[CASA:][]{mcmullin2007} for radio interferometric calibration and imaging. While the algorithm described by M19 is as presented in Section \ref{subsec:M19}, their implementation was different. This difference stems from the peculiarities of CASA, which were only discovered during our implementation of an improved version of AIRCARS, now christened P-AIRCARS \citep{Kansabanik2022_paircarsI}. AIRCARS uses the CASA task \textsf{tclean} for the imaging and deconvolution and produces a source model. To choose the baselines that were to be used for generating the source model, a list of antennas are passed to the \textsf{tclean}. \citet{Mondal2019} believed that CASA will only use the baselines between those antennas which were in the list. However, in this case, the implementation of CASA uses all baselines between antennas $p$ and $q$, where $p$ is the antenna in the list and $q$ is any other antennas where $p<q$. This implies that the implementation of M19 generated the starting model using all the baselines originating from the core. As will be discussed later, this ``error" made a significant contribution to producing the high dynamic range images that AIRCARS went on to produce.

As all core--core and core--noncore baselines are used in the calibration process from the beginning, first-order gain solutions of all antennas are available at every self-calibration round from the beginning. Baselines originating from antennas at larger distances are progressively added, which in turn add more constraints to the self-calibration problem and improve the gain solutions. When the phase solutions for all antennas are reasonably well constrained, amplitude-phase self-calibration is performed. Since the Sun is a source with very high flux density, any small error in gain amplitudes makes a large error in the amplitudes of the observed visibilities. Hence, the calibration of the instrumental gain amplitude makes a significant improvement in the DR after starting the amplitude phase self-calibration.
\begin{figure}
    \centering
    \includegraphics[trim={0cm 0cm 0cm 0cm},clip,scale=0.42]{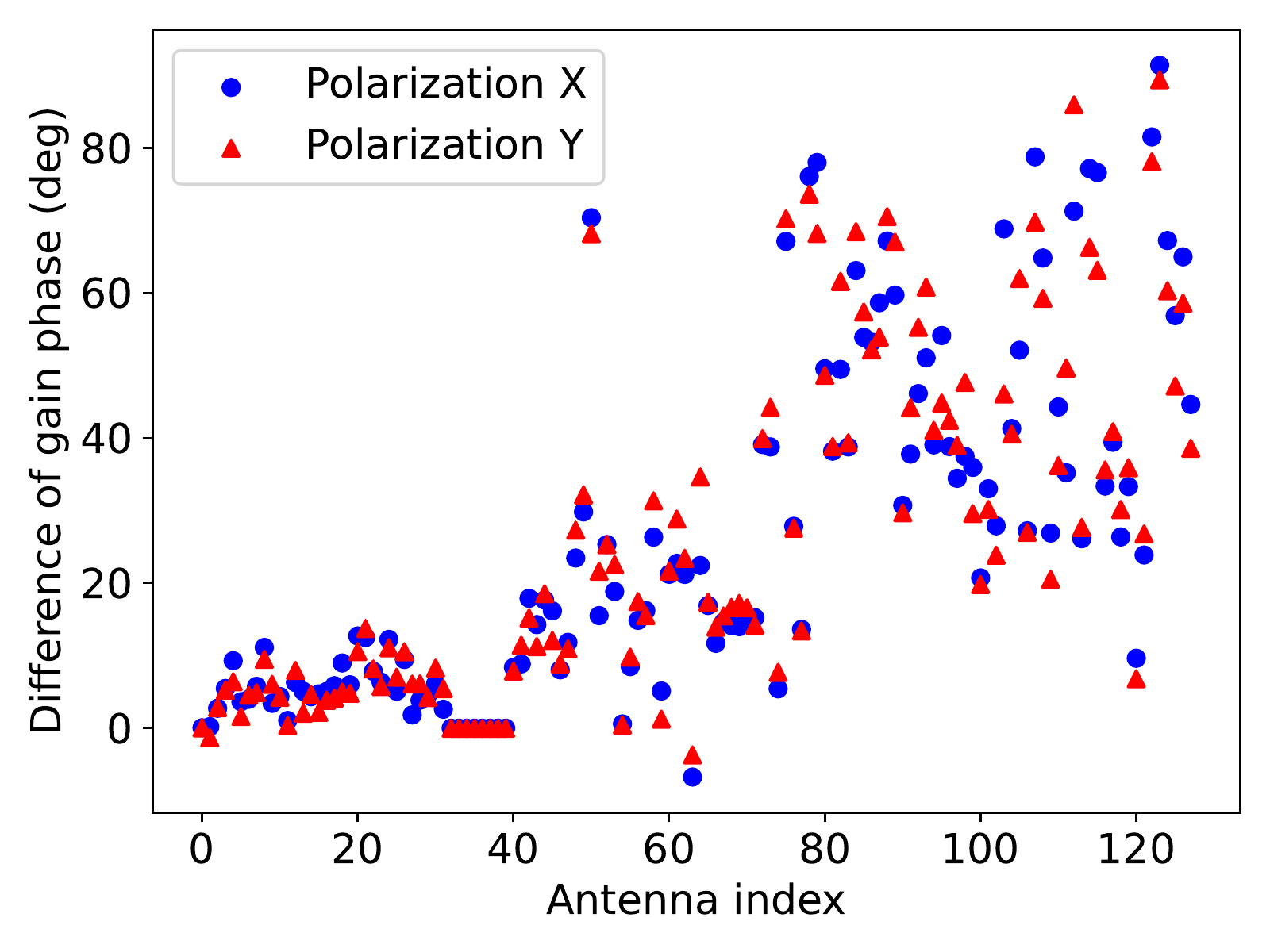}
    \caption{{\bf Difference between phases of the nighttime and daytime complex antenna gains.} \textit{Blue circle} and \textit{red triangle} represent the $X$- and $Y$-polarization, respectively.}
    \label{fig:cal_aircars_gain_diff}
\end{figure}

\begin{figure}
    \centering
    \includegraphics[trim={0cm 0cm 0cm 0cm},clip,scale=0.42]{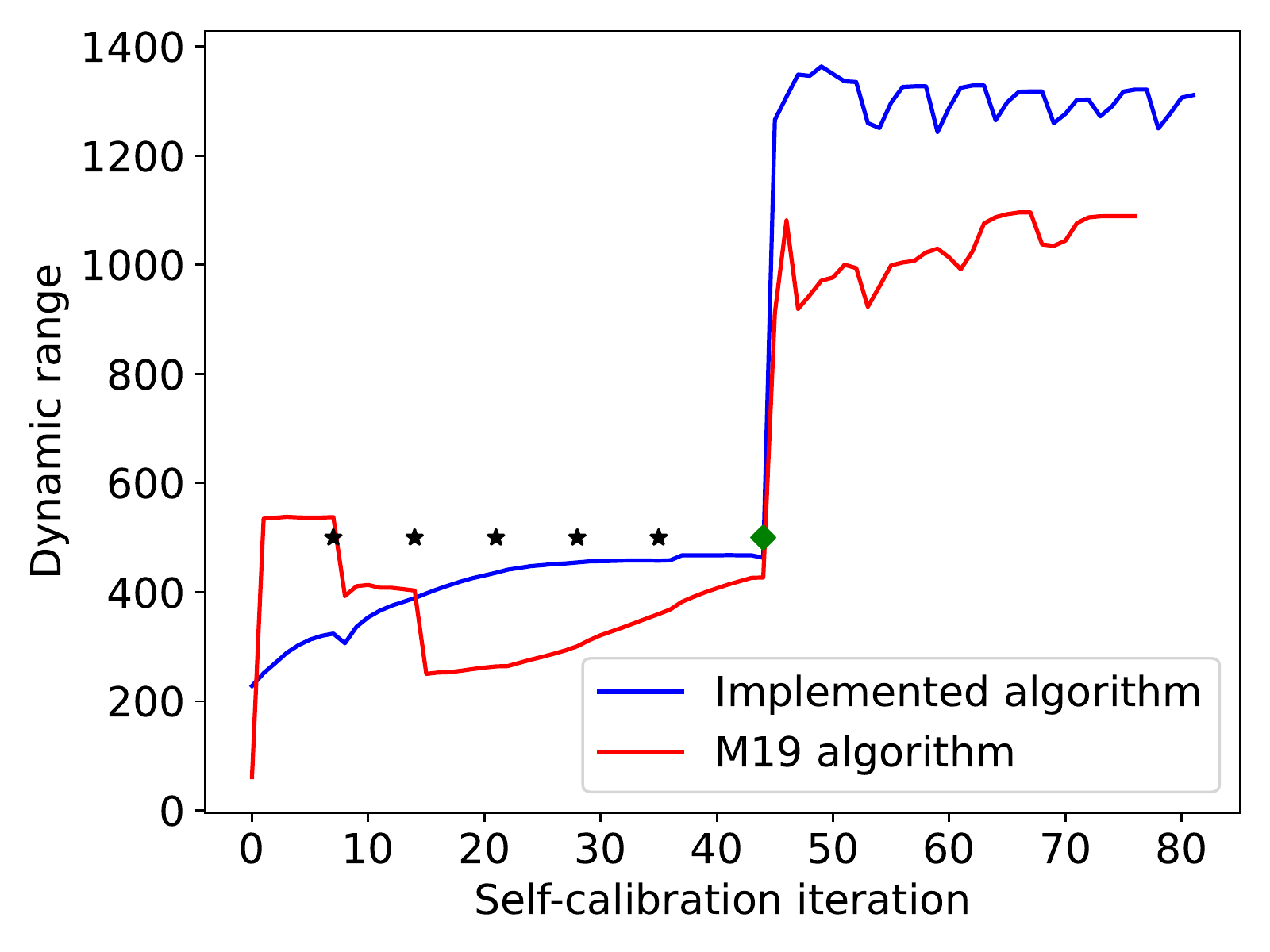}
    \caption{{\bf Comparison of the improvements in the dynamic range for M19 and implemented algorithm with the self-calibration iterations.} \textit{Black stars} show the iterations where new antennas are added  in the self-calibration process, while performing phase only self-calibrations. The \textit{green diamond} shows the iteration where amplitude-phase self-calibration starts with all antennas.}
    \label{fig:thought_vs_true}
\end{figure}

In the M19 algorithm, when new antennas are added to the calibration, the DR of the image suddenly drops and the source model also becomes worse, because the newly added antennas do not have any calibration. But, for the implemented algorithm, only a small number of baselines with increasing distance are added, and all antennas have a first-order gain solution. This introduces a much smaller error in the source model compared to adding a set of uncalibrated antennas. A comparison of the improvement in DR with self-calibration iterations between these two algorithms is shown in Figure \ref{fig:thought_vs_true}. It is evident from the figure that the DR improves almost monotonically for the implemented algorithm, while there are several drops for the M19 algorithm, which ends up with a smaller DR of the final image. In the case of the implemented algorithm, DR starts to oscillate after a few rounds of amplitude phase self-calibration. This implies the quicker convergence of the implemented algorithm compared to the M19 algorithm. In both cases, the DR of the final image is more than 1000. The decrease in DR by $\approx200-300$ may not affect the studies related to very bright radio bursts but is crucial when one tries to detect very weak emissions, like the gyrosynchrotron from CMEs \citep[][in preparation]{Mondal2020a,Kansabanik2022_GS} or the Weak Impulsive Narrow-band Quiet Sun Emissions \citep[WINQSEs:][]{Mondal2020b, Mondal2021b}. 

The rest of the article will focus on the implemented algorithm. One of the unique features of the AIRCARS is that it can start the self-calibration even without any a-priori calibration solutions obtained from the night-time calibrators. In the later sections, the explanation behind this unique feature of AIRCARS is discussed in detail.

\section{Initial Source Model of AIRCARS}
\begin{figure}
    \centering
    \includegraphics[trim={3cm 14cm 3cm 0cm},clip,scale=0.4]{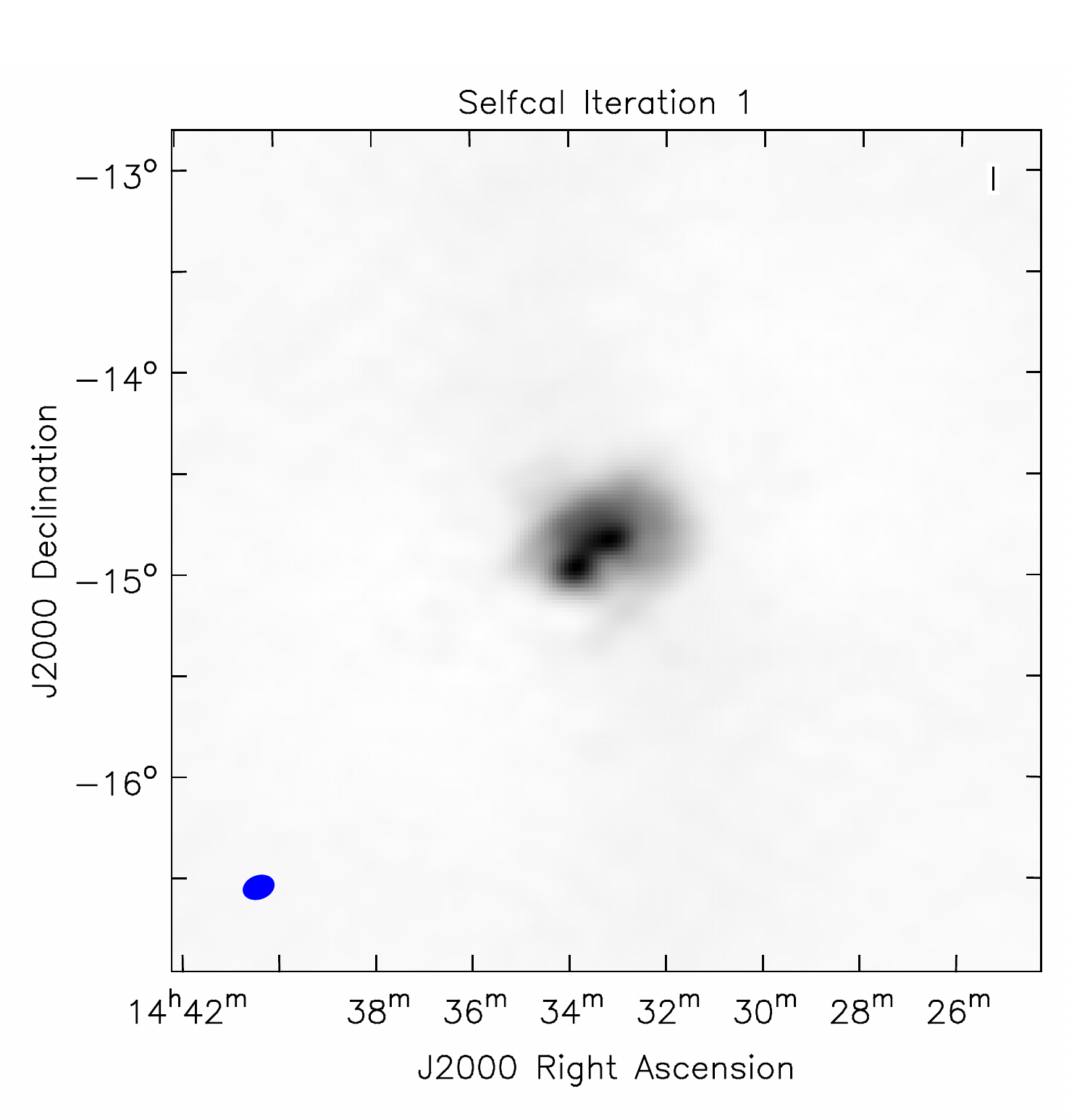}\includegraphics[trim={3cm 14cm 3cm 0cm},clip,scale=0.4]{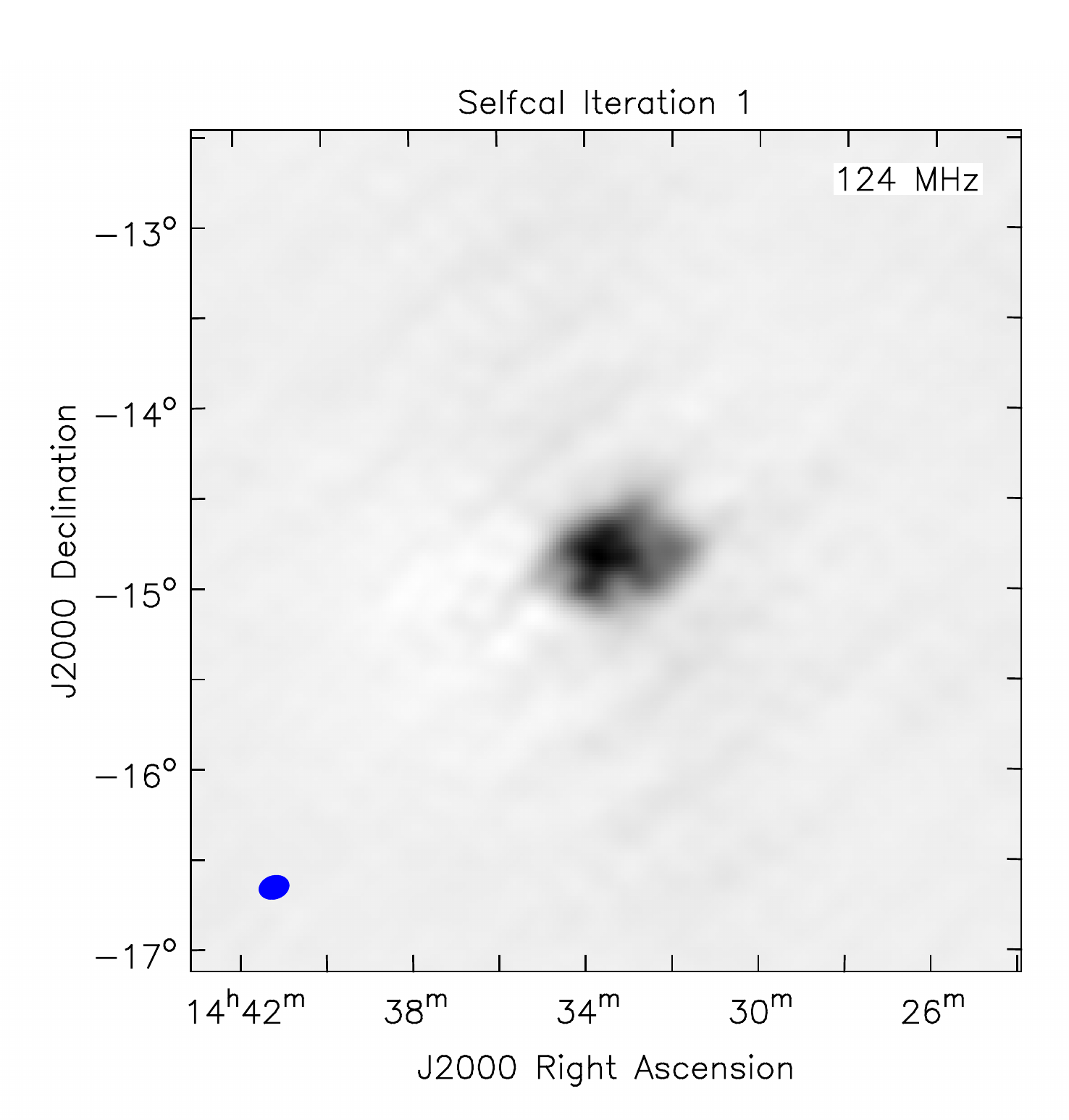}
    \caption{{\bf Initial image of the Sun.} \textit{Left-panel : }Initial image made after applying the calibration solutions from the nighttime calibrator observation. \textit{Right-panel : }Initial image made directly from the uncalibrated visibilities which have at least one core antenna. The {\it filled  blue ellipses} at the bottom left of the figures represent the PSF of the array.}
    \label{fig:ini_model}
\end{figure}

When nighttime calibration is available, the calibration solutions are applied first, and then the initial image is made (left panel of Figure \ref{fig:ini_model}). When this is not available, the initial image is made from the uncalibrated observed visibilities (right panel of Figure \ref{fig:ini_model}). There is no drastic difference between these two images. This happens because the phases of the gains during daytime are significantly different from the phases during nighttime (Figure \ref{fig:cal_aircars_gain_diff}), hence the nighttime calibration solutions may not always produce any noticeable correction to the observed visibilities. In both cases, there is a significant amount of source flux concentrated near the phase center, because the phase distribution of the antenna gains is not uniformly random, which is shown later in Sections \ref{sec:expected_gain_char} and \ref{sec:stats}. 

\section{Expected Characteristics of the Complex Gains}\label{sec:expected_gain_char}
As described in Sec. \ref{sec:calibration}, $J_\mathrm{p}(\nu,\ t,\ {\bf l})$ can be decomposed into $G_\mathrm{p}(t), B_\mathrm{p}(\nu)$, and $E_\mathrm{p}({\bf l})$. As discussed in Section \ref{sec:direction_dependent_calibration}), $E_\mathrm{p}({\bf l})$ can be neglected for solar observation. The algorithm determining $B_\mathrm{p}(\nu)$ is out of the scope of the article and has already been described in detail at \citet{Kansabanik2022,Kansabanik2022_paircarsI}. The only remaining term is time-dependent complex gain: $G_\mathrm{p}(t)$. $G_\mathrm{p}(t)$ has contributions from the both instrument [$g_\mathrm{p}^{\mathrm{instrumental}}(t)$] and the ionosphere [$g_\mathrm{p}^{\mathrm{ion}}(t)$]. In practice, it is not necessary to separate them and is not done in AIRCARS. 

\subsection{Expected Characteristics of the Instrumental Gains}\label{subsec:instrumental_gain}
The contributions from $g_\mathrm{p}^{\mathrm{instrumental}}(t)$ are not expected to originate from a uniform random distribution. There are several reasons behind this :
\begin{enumerate}[label=\roman*)]
    \item Except for the active dipoles and the low-noise amplifiers (LNA), other components of the electronic chain of the MWA are passive elements \citep{Tingay2013}, and the characteristics of the passive components are extremely stable.
    \item The characteristics of the LNA can also be well modeled \citep{Sokowlski2017} for the MWA, and they are similar for all the antenna elements.
    \item Temperature variation of the environment changes the cable length and introduces an additional phase to the complex gain. These are small for the core antennas, which are connected using small cables, and, grow larger for the antennas at long baselines connected using longer cables.
    \item Despite the well-modeled LNA and passive elements, there are some manufacturing tolerances, which could introduce a spread in the distribution of the instrumental gains.
\end{enumerate}

\subsection{Expected Characteristics of the Ionospheric Phases}\label{subsec:ionopsheric_phase_expectations}
At low radio frequencies, another major contribution to the complex gain comes from the ionosphere. \citet{mondal_ionosphere} determined the total electron content (TEC) using daytime observation of the Sun along a single line of sight. They demonstrated that the daytime ionospheric TEC can vary over the MWA array, even over the core. The TEC value varies by $\approx10\ \mathrm{mTECU}$ over the core antennas, which corresponds to $\approx50\ \mathrm{degrees}$ \citep{Mevius2015} (left panel of Figure 1 of \citet{mondal_ionosphere}). \citet{mondal_ionosphere} also showed that the variation is smooth across the array, and the mean subtracted small-scale random TEC fluctuations over the array is $\lesssim1\ \mathrm{mTECU}$ (middle panel of Figure 1 of \citet{mondal_ionosphere}), which corresponds to a few degrees \citep{Mevius2015} of ionospheric phase variations. This demonstrates that although there are variations of the ionosphere across the MWA array, and even over the core antennas, this variation is smooth and the random fluctuations are small. 

\subsection{Expected Statistical Properties of $G_\mathrm{p}$} \label{subsec:expected_stats}
As described in Sections \ref{subsec:instrumental_gain} and \ref{subsec:ionopsheric_phase_expectations}, the core antennas are expected to have a similar phase with a spread from a mean value due to instrumental (temperature variation across the array, manufacturing tolerances) and ionospheric effects. These effects become large away from the core. Hence one can expect the following distribution of the phases of $G_\mathrm{p}$ :
\begin{enumerate}[label=\roman*)]
    \item {\bf Only core antennas}: Distribution will be quasi-Gaussian with a small standard deviation.
    \item {\bf Only non-core antennas}: Distribution will not be a peaked distribution and the standard deviation will be very large.
    \item {\bf All antennas}: Since the core antennas ($\approx60$) dominate the total number of antennas, distribution will be quasi-Gaussian with a slightly larger standard deviation compared to the distribution of only core antennas.
\end{enumerate}

\begin{figure}
    \centering
    \includegraphics[trim={0.5cm 0cm 0cm 0cm},clip,scale=0.16]{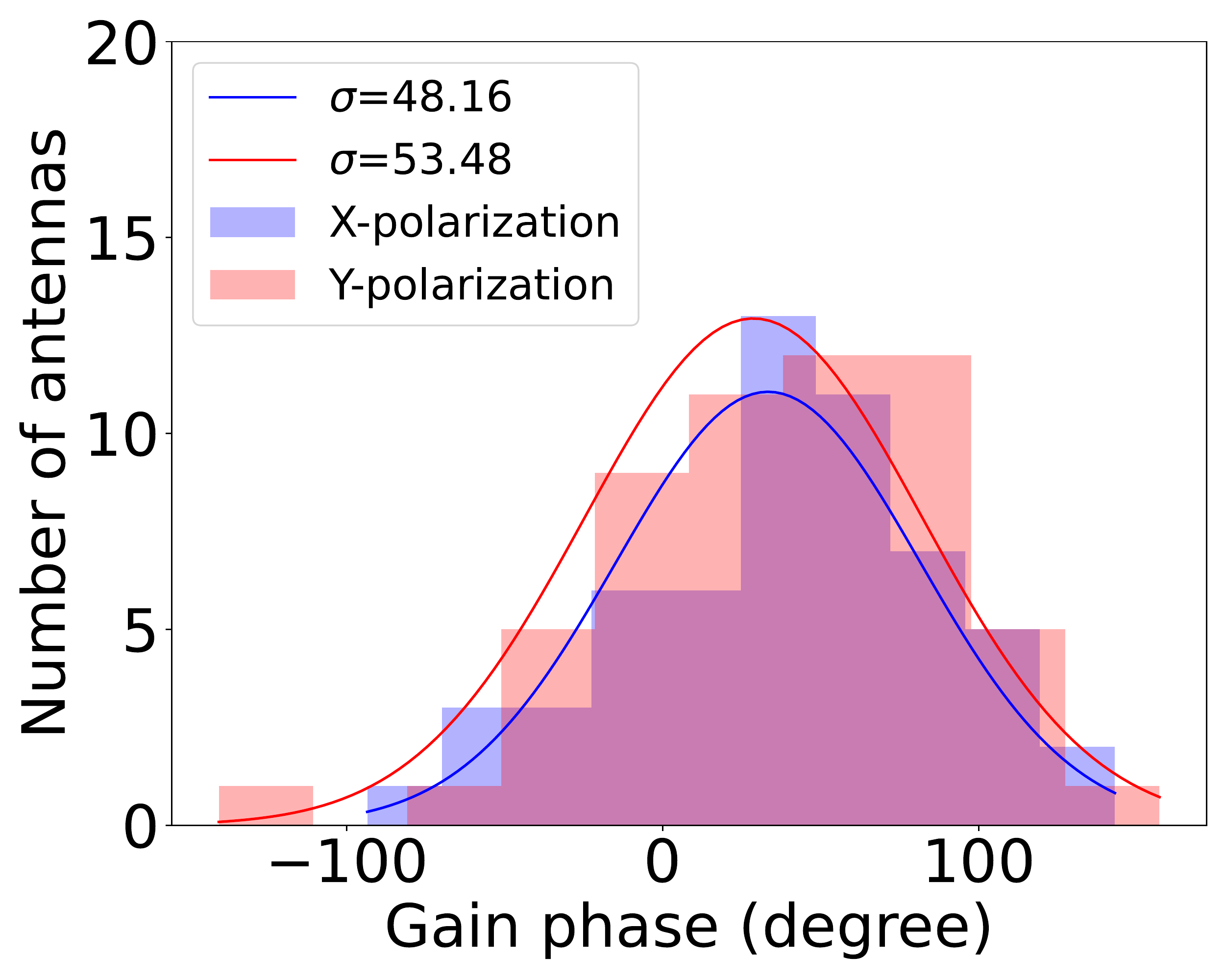}
    \includegraphics[trim={1.7cm 0cm 0cm 0cm},clip,scale=0.16]{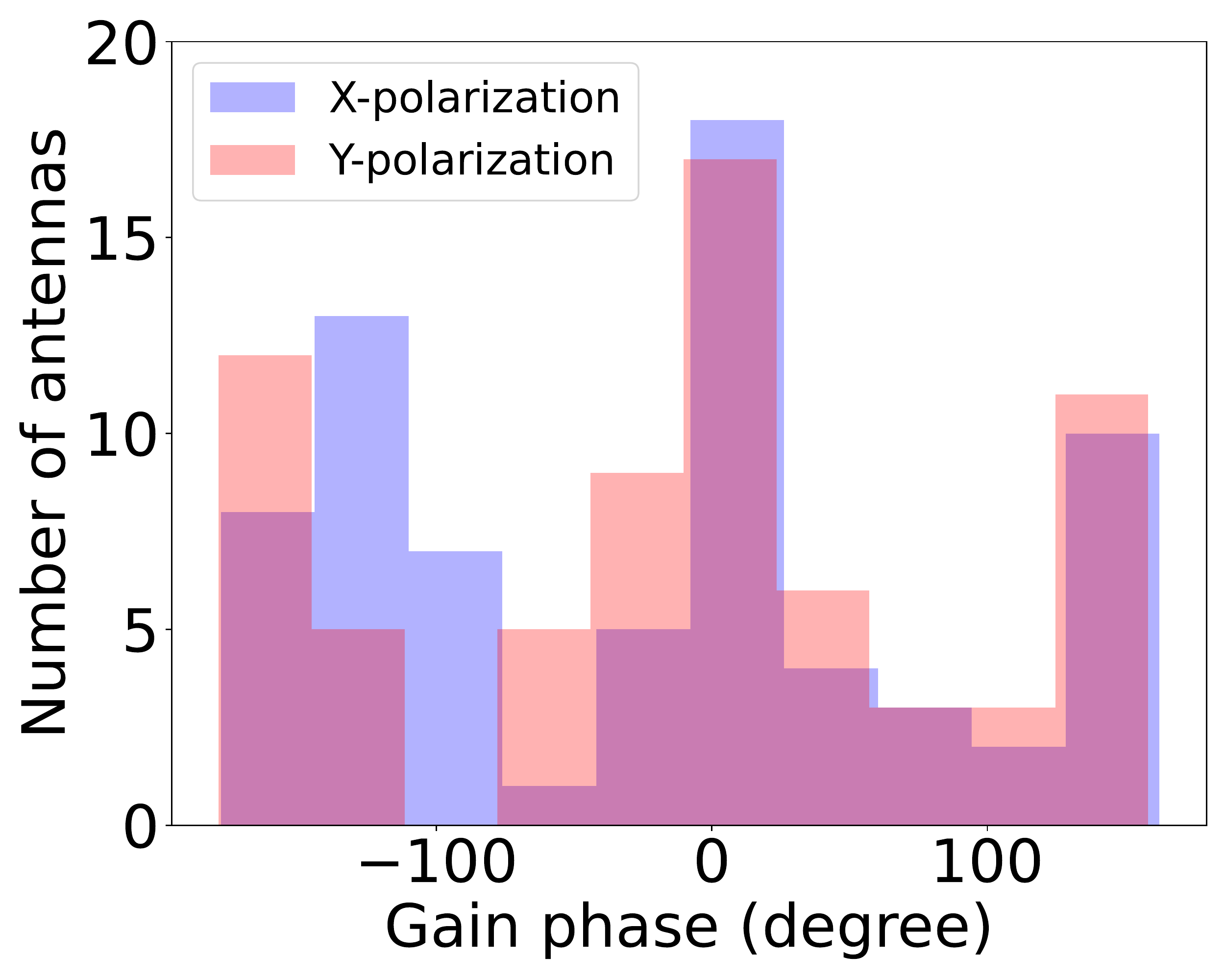}
    \includegraphics[trim={1.7cm 0cm 0cm 0cm},clip,scale=0.16]{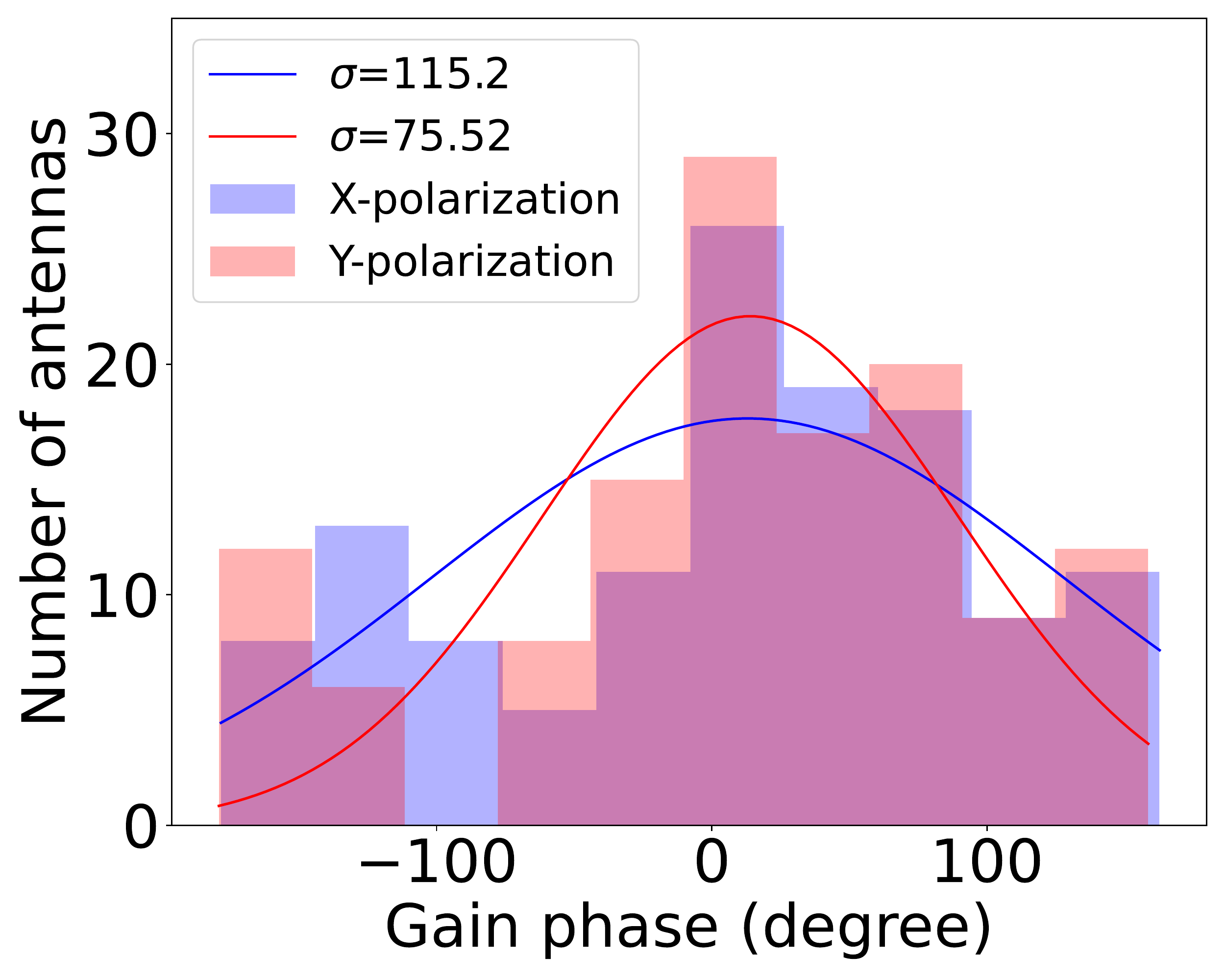}
    \caption{{\bf Distribution of phases of the antenna gains at 80 MHz for the observation on 05 May 2014.} $X$-polarization is shown by {\it blue} and $Y$-polarization is shown by {\it red}. Distribution of phases of $G_\mathrm{p}$ is shown for \textit{Left-panel : } only the core antennas, \textit{Middle-panel : }for the non-core antennas, and \textit{Right-panel : }for all antennas.}
    \label{fig:gain_stats}
\end{figure}

\section{Comparison Between the Expected and Observed Statistical Properties of the Antenna Gains}\label{sec:stats}
A comparison between the expected and observed properties is done for the three sub-groups of antennas as mentioned in Section \ref{subsec:expected_stats}. The coherency of $V^\prime_{pq}$ depends on the coherency of $J_\mathrm{p}$. Although the coherency depends on both the amplitude and phase of $J_\mathrm{p}$, it is well known that a small loss in phase coherency causes a significant loss in coherency loss of $V^\prime_{pq}$ compared to amplitudes \citep{thompson2017}. It is evident from Equation \ref{eq:measurement_eq} that the coherence of $V^\prime_{pq}$ is affected by the phase difference: $\phi_p-\phi_q$. Hence, the statistical properties of the $\phi_p-\phi_q$ are also discussed.

\subsection{Observed Properties of Antenna Gains}\label{subsubsec:antenna_gain}
The histograms of the phases of the complex gains are shown in Figure \ref{fig:gain_stats}. The distribution of phases for only the ``core antennas" is shown in the left panel and well fitted with a Gaussian distribution with a standard deviation of $\approx50\ \mathrm{degrees}$. The distribution of the phases of only for the ``non-core antennas" is shown in the middle panel, and could not be fitted with a Gaussian distribution. The distribution of phase for ``all antennas" is shown in the right panel. Although a Gaussian has been fitted, the fitting is not good. The standard deviation of the distribution is $\approx100$ degrees, which is large compared to the ``core antennas". These observed properties match the expected properties as mentioned in Section \ref{subsec:expected_stats}.

\begin{figure}
    \centering
    \includegraphics[scale=0.24]{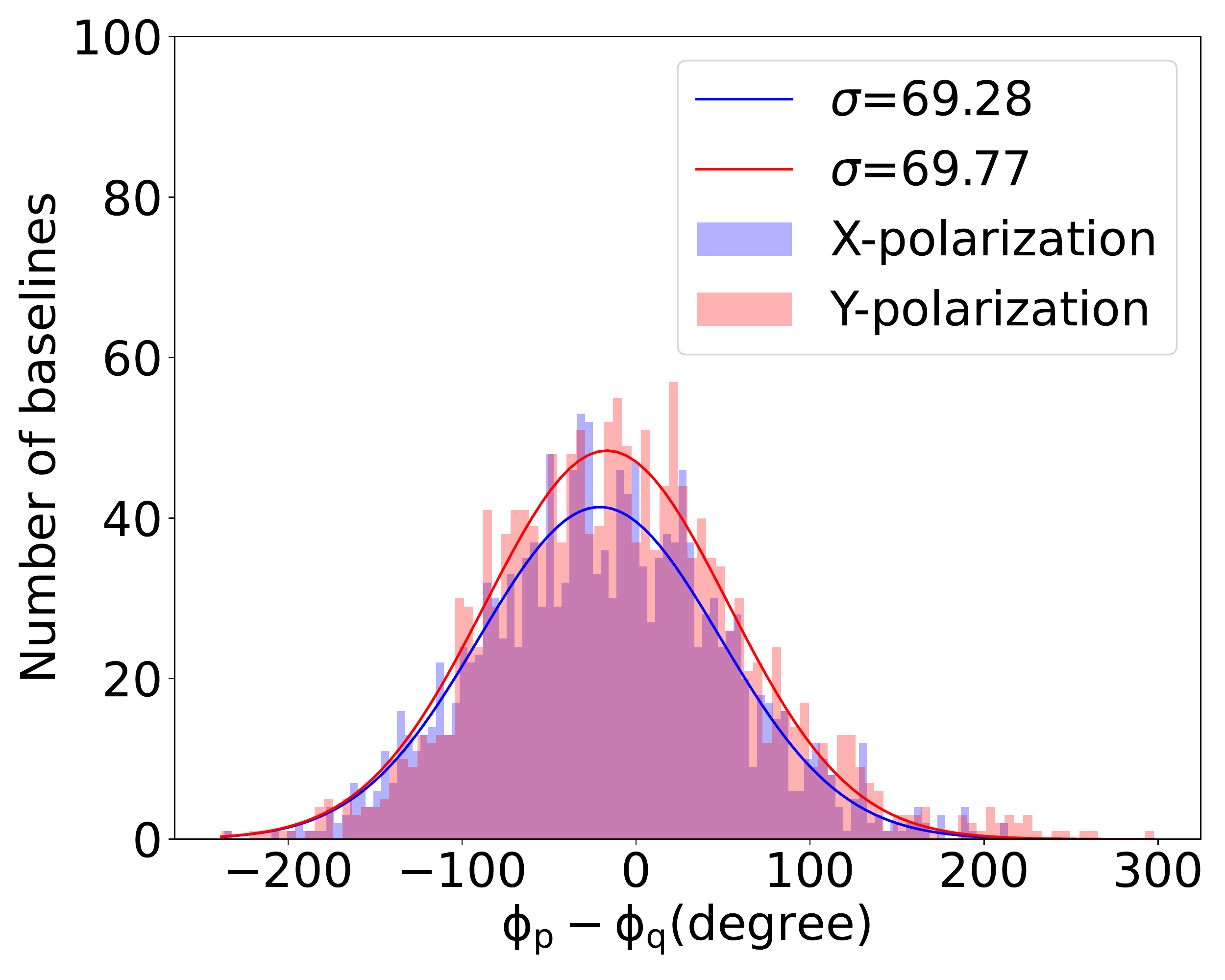}\includegraphics[scale=0.24]{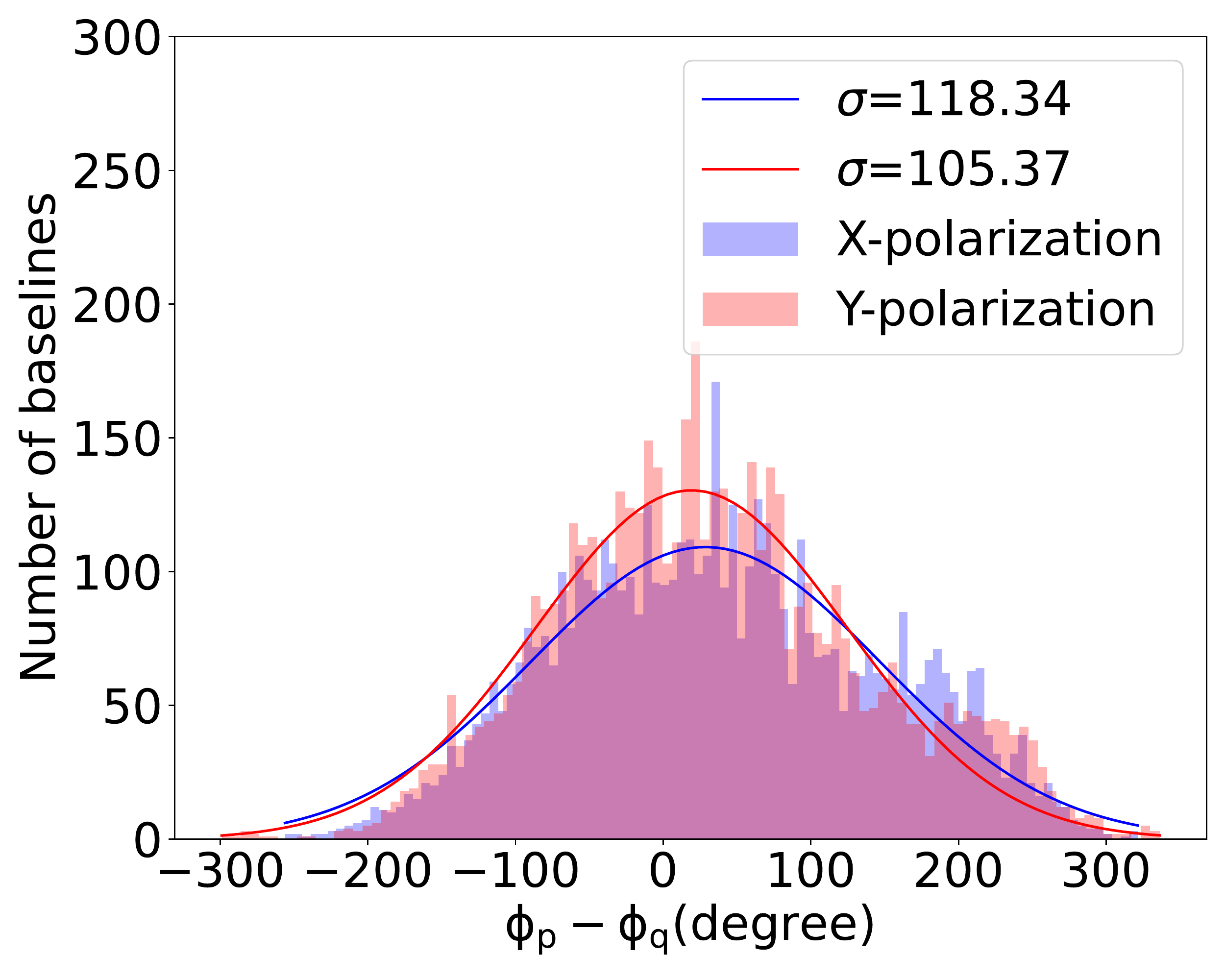}
        \caption{{\bf Distribution of $\phi_\mathrm{p}-\phi_\mathrm{q}$ at 80 MHz for the observation on 05 May 2014. X polarization is shown by blue and Y polarization is shown by red. Left panel: }Distribution for core-all baselines. {\bf Right panel: }Distribution of all baselines.}
    \label{fig:baseline_stats}
\end{figure}

\subsection{Observed Properties of $\phi_\mathrm{p}-\phi_\mathrm{q}$}
The histogram of $\phi_\mathrm{p}-\phi_\mathrm{q}$ for all the baselines originating from the core (core-all) is shown at the left panel and for all baselines is shown at the right panel of Figure \ref{fig:baseline_stats}. The standard deviation of the fitted Gaussian for the core-all histogram is much smaller ($\approx70\ \mathrm{degrees}$) compared to all baselines ($\approx120\ \mathrm{degrees}$). Both these distributions follow a Gaussian distribution but, there are still slight deviations from the true Gaussian distribution at the edges, which is more prominent for all baselines.  

The observed statistical properties of both the phases and the difference between the phases of the antenna gains follow a quasi-Gaussian distribution. It does not follow a ``uniform random" distribution. For all baselines, the standard deviation of the Gaussian becomes larger and also starts to deviate from the true Gaussian distribution, and the array loses coherency. However, this standard deviation is much smaller for core-all baselines, which provides better coherency to the uncalibrated observed visibilities. Availability of this starting source model (Figure \ref{fig:ini_model}) is the reason why AIRCARS can produce high dynamic range images through self-calibration approach alone.

\section{Simulation}\label{sec:simulation}
In Section. \ref{sec:stats}, it is stated that AIRCARS can proceed with the self-calibration from the uncalibrated observed visibilities because the distribution of phase and phase difference is not uniform random. This phenomenological explanation is verified through simulation in this section.

\begin{figure}
    \centering
    \includegraphics[trim={0cm 14cm 0cm 0cm},clip,scale=0.42]{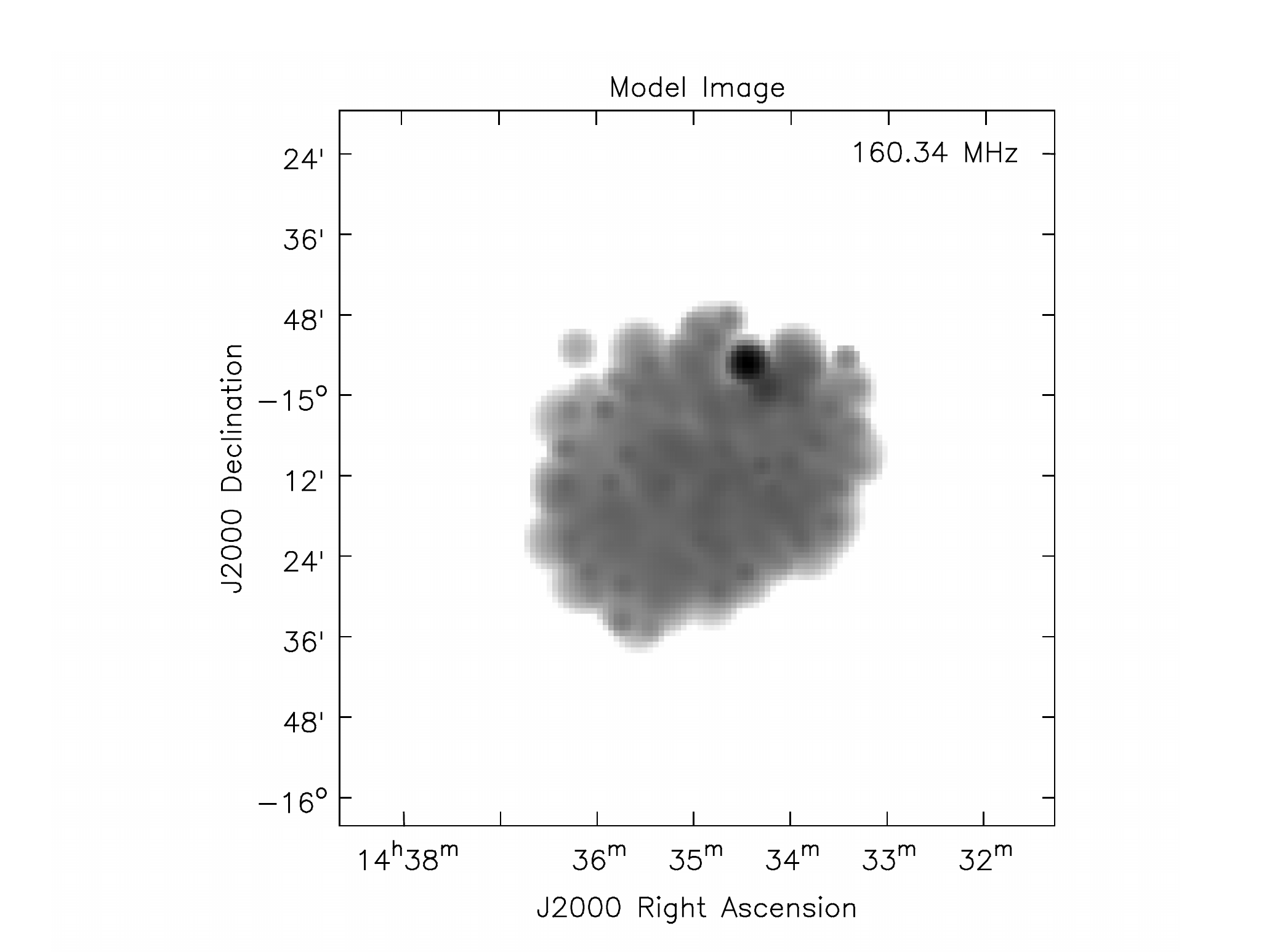}
    \caption{Model image of the Sun from the observation on 11 November 2015 used for simulation.}
    \label{fig:model}
\end{figure}

\subsection{Description of the simulation}
\begin{figure}
    \centering
    \includegraphics[trim={0cm 0cm 0cm 0cm},clip,scale=0.29]{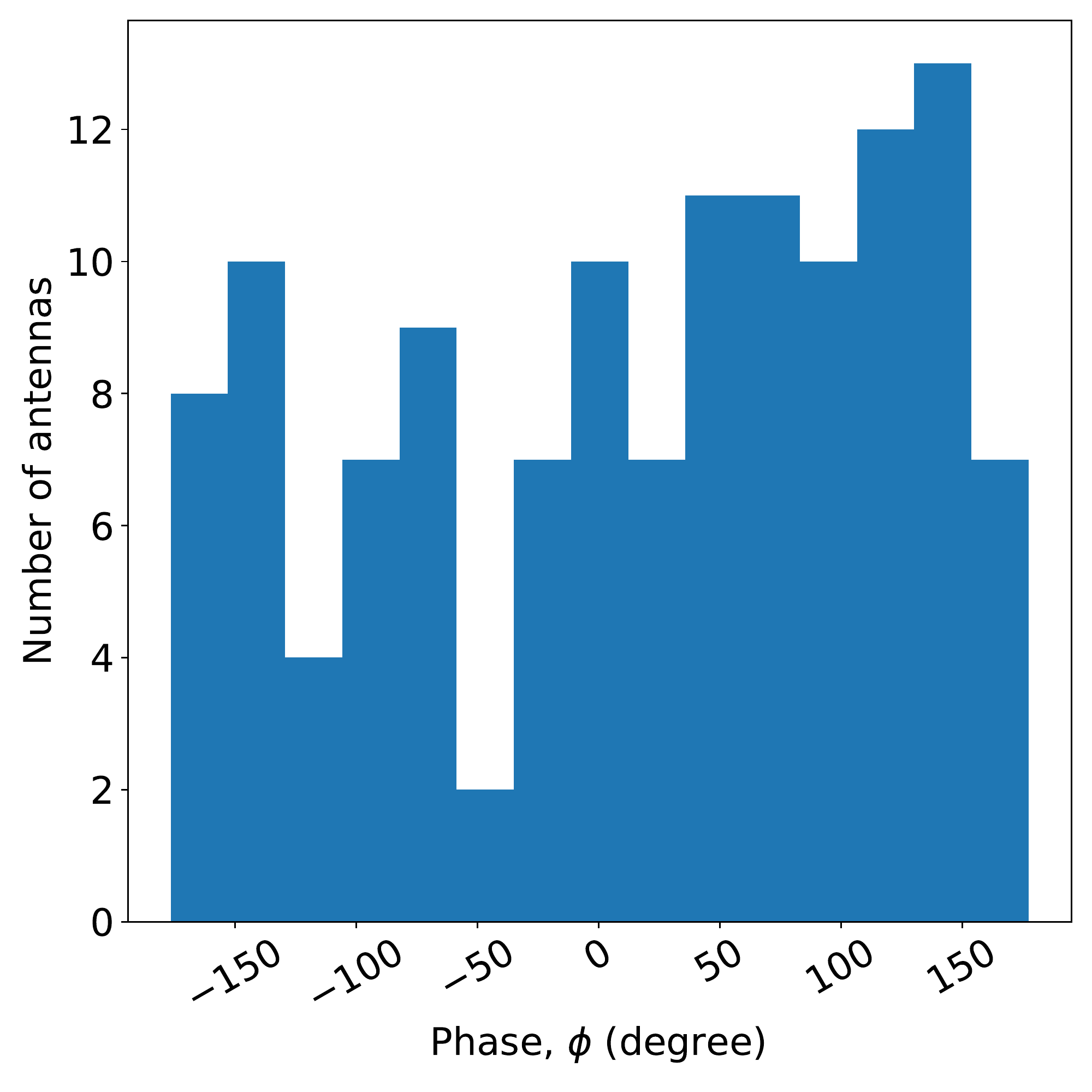}\includegraphics[trim={3cm 14cm 3cm 1cm},clip,scale=0.42]{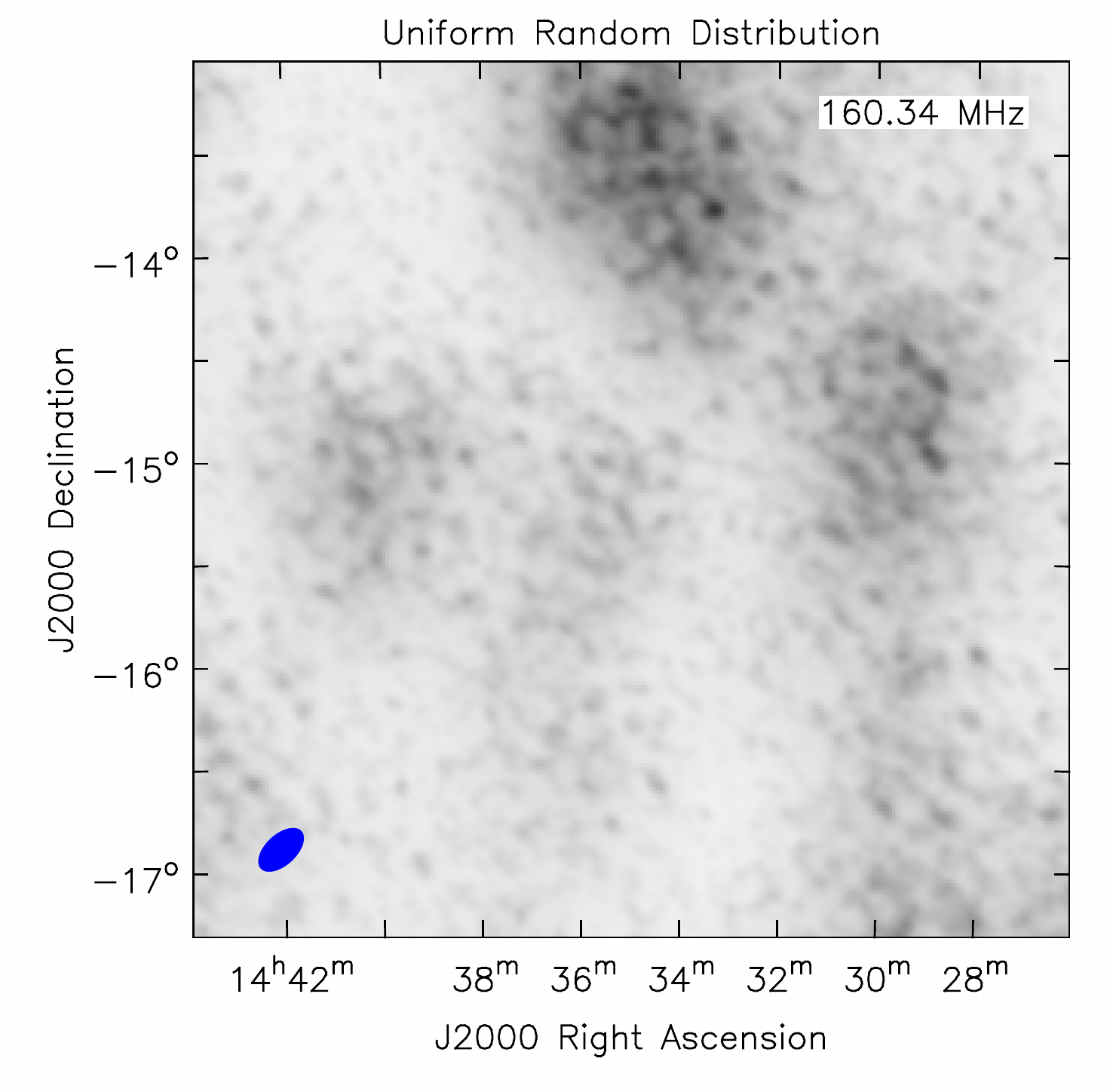}
    \caption{\textit{Left-panel : }Uniform random distribution of the phase of the antenna gains. \textit{Right-panel : } Dirty image made from simulated visibilities. The filled {\it blue ellipse} at the bottom represents the PSF of the array.}
    \label{fig:uniform}
\end{figure}

The simulation is done as follows :
\begin{enumerate}[label=\roman*)]
    \item A model image of the Sun is obtained from the observation on 11 November 2015 (Figure \ref{fig:model}). This model is obtained using the {\bf imaging and deconvolution} task \textsf{tclean} of the CASA. A number of Gaussians having multiple sizes \citep{cornwell2008} are used in the deconvolution process to produce this model.
    \item The model image is then Fourier transformed to the model {\it visibilities}: $V_\mathrm{{pq,model}}$.
    \item Antenna gains [$G_\mathrm{p}$] are simulated from a underlying distribution. Amplitudes are chosen to be unity. 
    \item Simulated visibilities are obtained as, $V_\mathrm{{pq}}^\prime=G_\mathrm{p}\ V_{\mathrm{pq,model}}\ G^\dagger_\mathrm{q}$. 
\end{enumerate}
The antenna gains are drawn from the two types of distributions between $-180$ and $+180\ \mathrm{degrees}$ : 
\begin{enumerate}[label=\roman*)]
    \item {\bf Uniform random distribution : }The probability density function of the uniform random distribution is given as
    \begin{equation}
        p(x; \mathrm{a, b})=\frac{1}{\mathrm{a-b}}
    \end{equation}
     within the interval $\mathrm{[a, b)}$, and zero outside this range. 
     \item {\bf Truncated Gaussian random distribution: }The probability distribution function is given as
     \begin{equation}
     \begin{split}
        p(x;\mu,\sigma,\mathrm{a,b)}=\frac{\phi({\frac{x-\mu}{\sigma}})}{\Phi(\mathrm{\frac{b-\mu}{\sigma}})-\Phi(\mathrm{\frac{a-\mu}{\sigma}})}
     \end{split}
     \end{equation}
     for $\mathrm{a}\leq x\leq \mathrm{b}$ and $p=0$ otherwise. Here, $\phi(\zeta)$ is the probability distribution function of standard Gaussian distribution:
     \begin{equation}
     \begin{split}
         \phi(\zeta)=\frac{1}{\sqrt{2\pi}}\mathrm{exp(-\frac{1}{2}\zeta^2)}
     \end{split}
     \end{equation}
     and, $\Phi(\epsilon)=\frac{1}{2}[1+\mathrm{erf(\frac{\epsilon}{\sqrt{2}})}]$ is the cumulative distribution function.
\end{enumerate}

\begin{figure}
    \centering
     \includegraphics[trim={0cm 0cm 0cm 0cm},clip,scale=0.29]{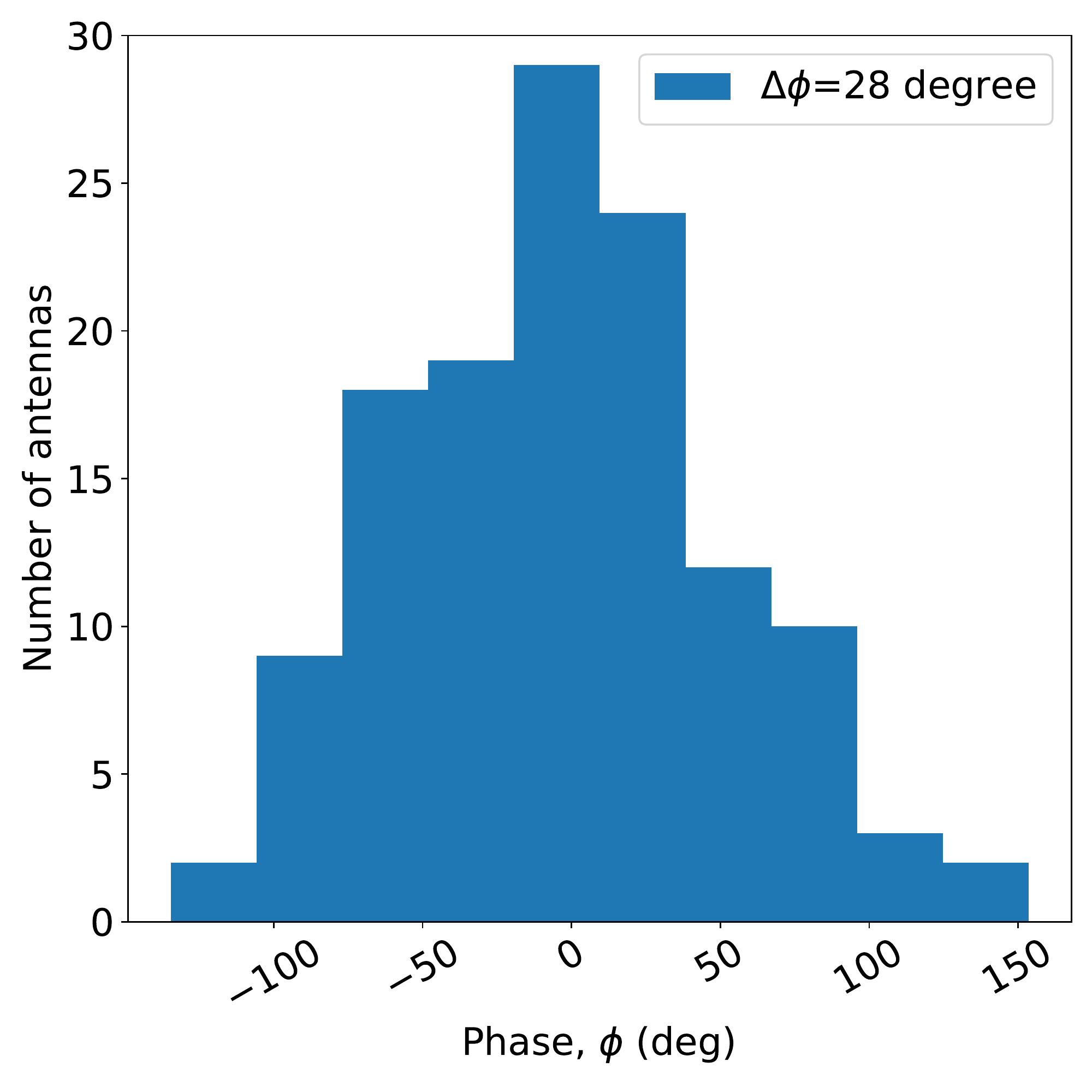}\includegraphics[trim={3cm 14cm 3cm 1cm},clip,scale=0.42]{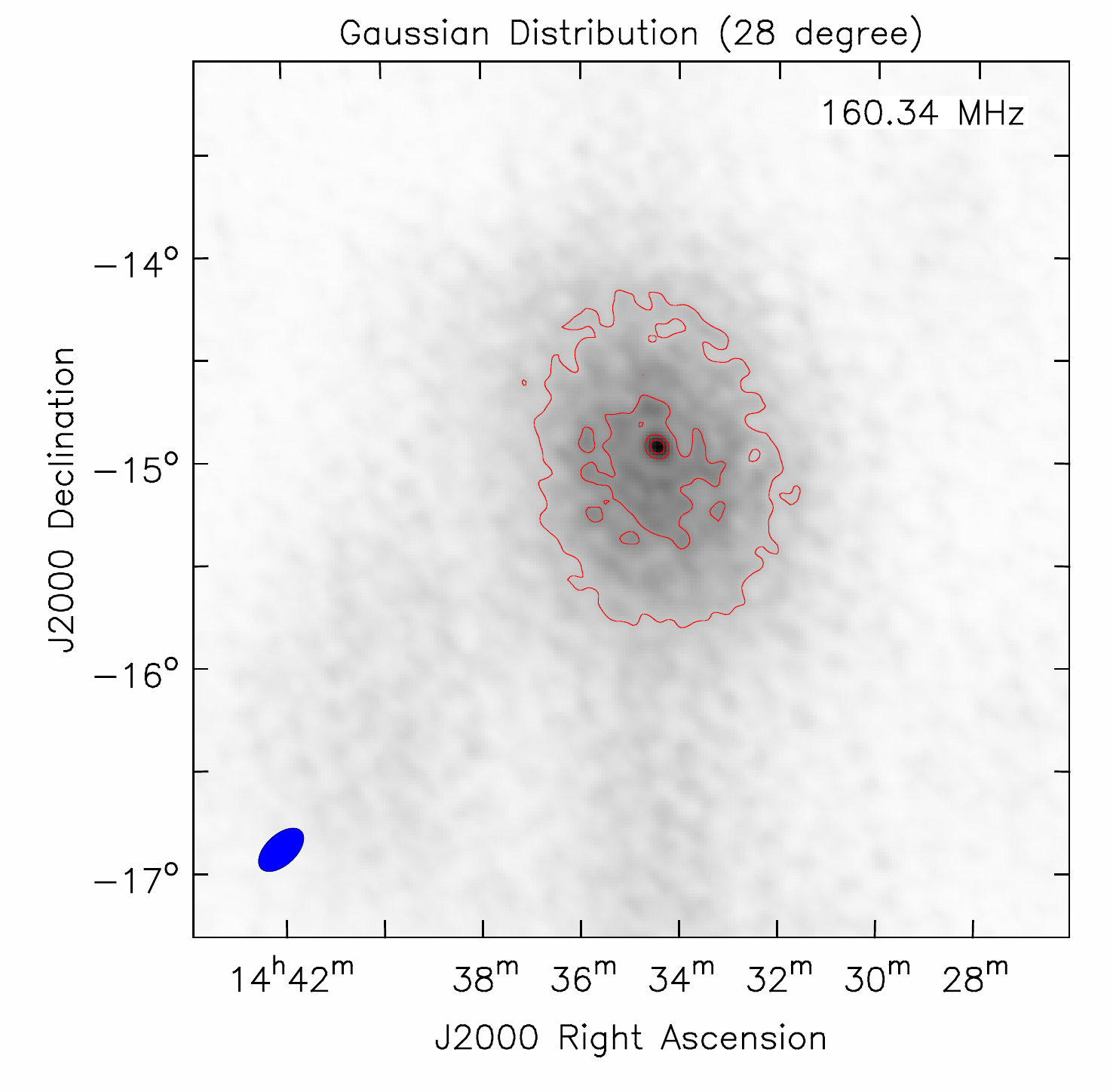}\\
    \includegraphics[trim={0cm 0cm 0cm 0cm},clip,scale=0.29]{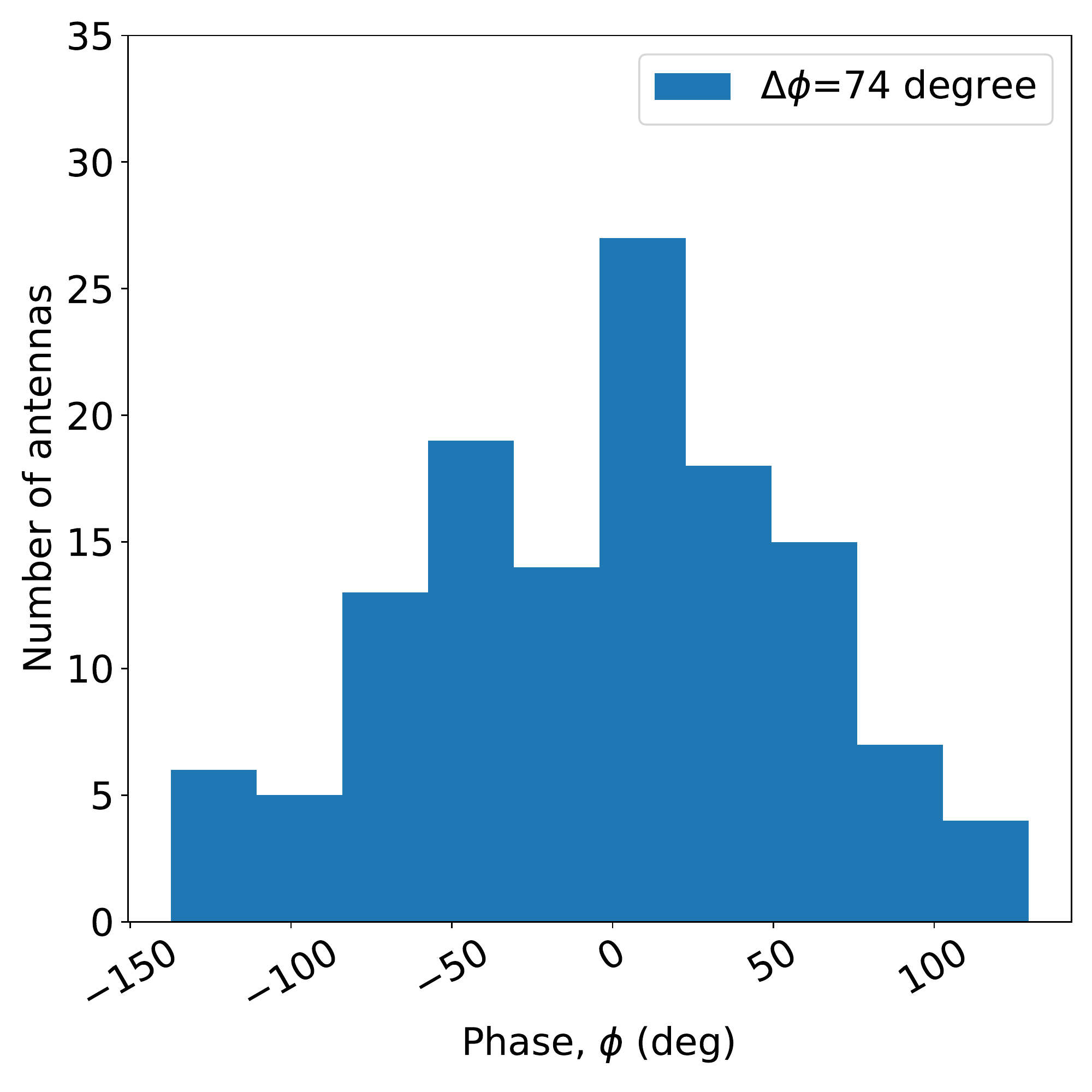}\includegraphics[trim={3cm 14cm 3cm 1cm},clip,scale=0.42]{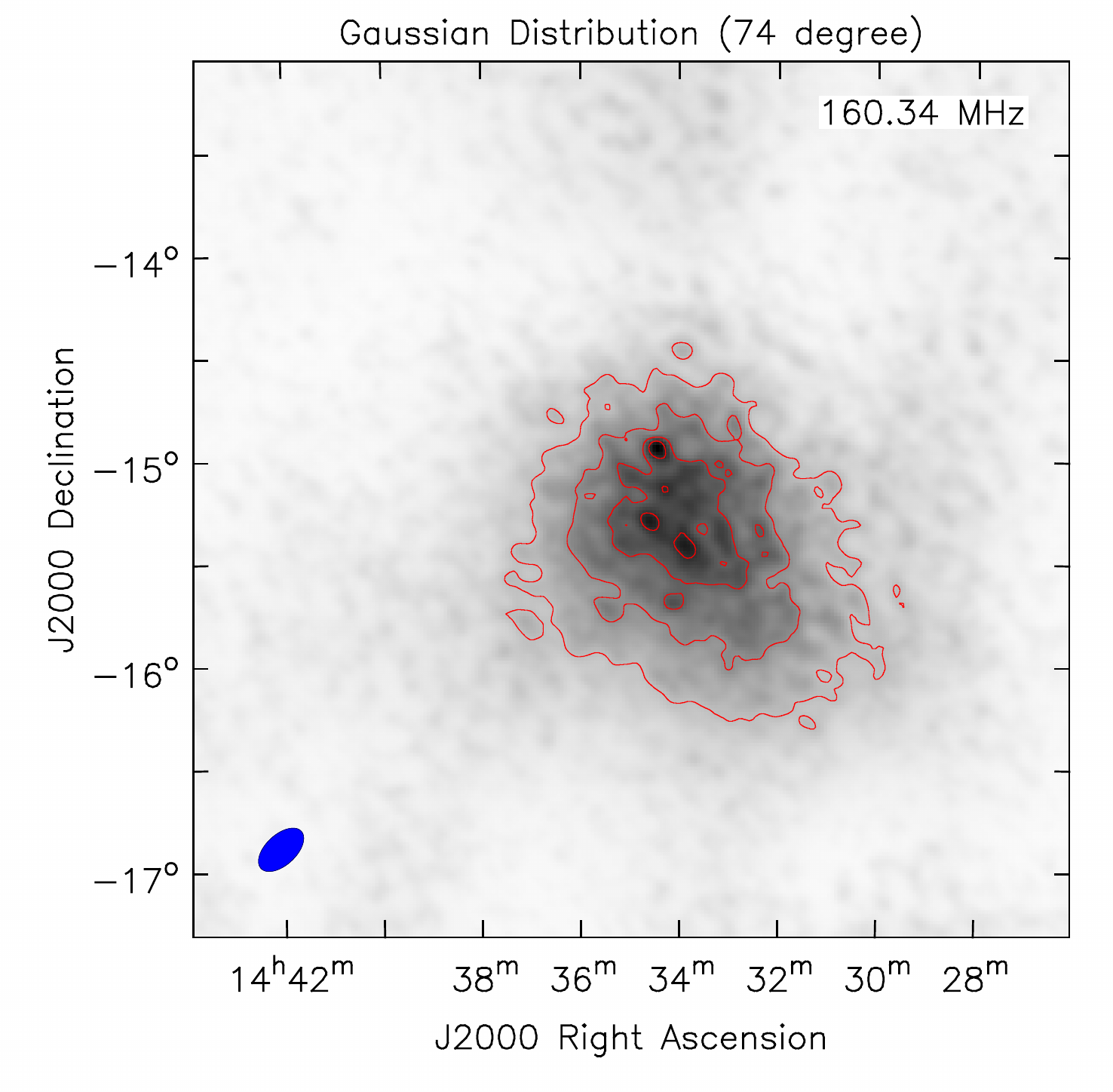}\\
    \includegraphics[trim={0cm 0cm 0cm 0cm},clip,scale=0.29]{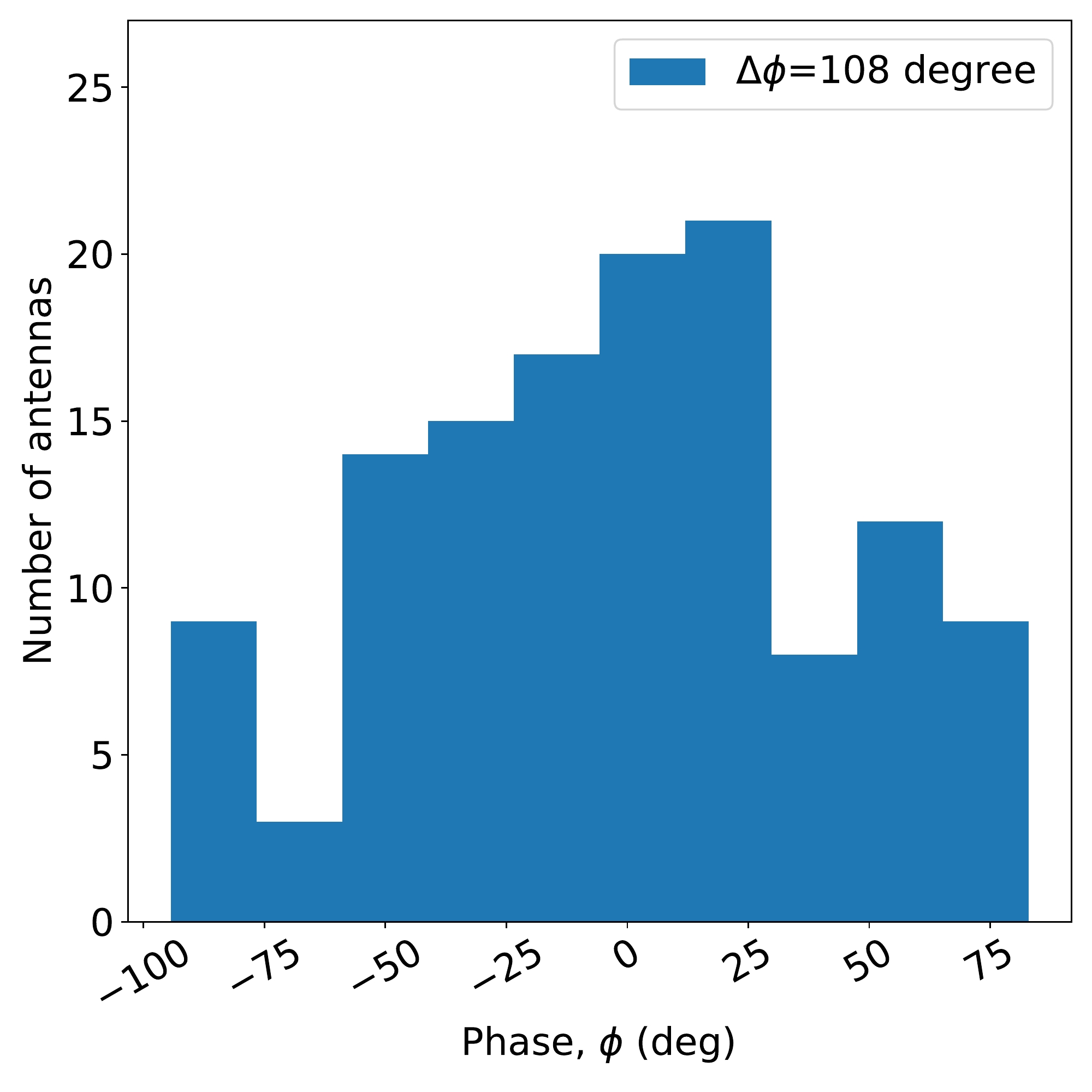}\includegraphics[trim={3cm 14cm 3cm 1cm},clip,scale=0.42]{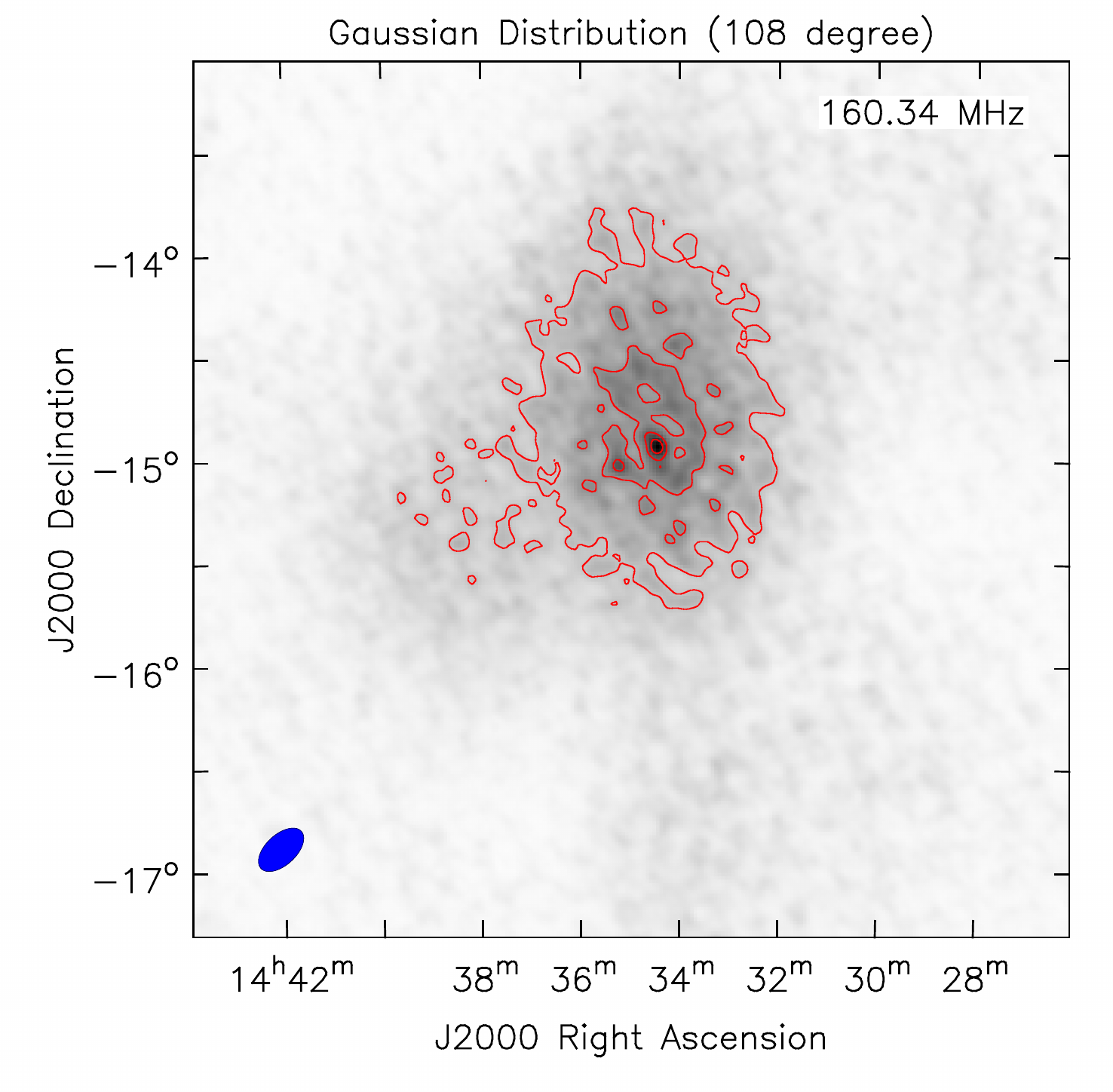}
    \caption{{\bf Dirty images from simulated visibilities from truncated Gaussian phase distribution of antenna gains.} {\it Left panels} show the distribution of the simulated phases and the {\it right panels} show the corresponding dirty images. Results are shown for three truncated Gaussian distributions with standard deviations; \textit{Top panel : }28 degrees. \textit{Middle panel : } 74 degrees, and \textit{Bottom panel : }108 degrees. The ﬁlled blue ellipses at the bottom left of the right panels represent the PSF of the array.} 
    \label{fig:gaussian}
\end{figure}

\subsection{Properties of the Initial Images Made from Simulated Visibilities}
The prime goal of this simulation is to demonstrate the suitable distribution of the phase, such that uncalibrated visibilities have some coherency and AIRCARS can start the calibration without any dedicated calibrator observations.

The dirty image made from the simulated visibilities for a uniform random distribution of the phase of the antenna gains is shown in Figure \ref{fig:uniform}. There is no source detected with more than 10-$\sigma$ (default value used in AIRCARS to pickup emission in the source model) significance near the phase center and the image looks noise-like. This demonstrates that if the phases of the antenna gains follow a uniform random distribution, the array does not have any coherency. Unlike a Gaussian distribution of phases, in this case, it is not possible to start the self-calibration without any dedicated calibrator observation.

\begin{figure}
    \centering
    \includegraphics[trim={0cm 0.2cm 0cm 0cm},clip,scale=0.42]{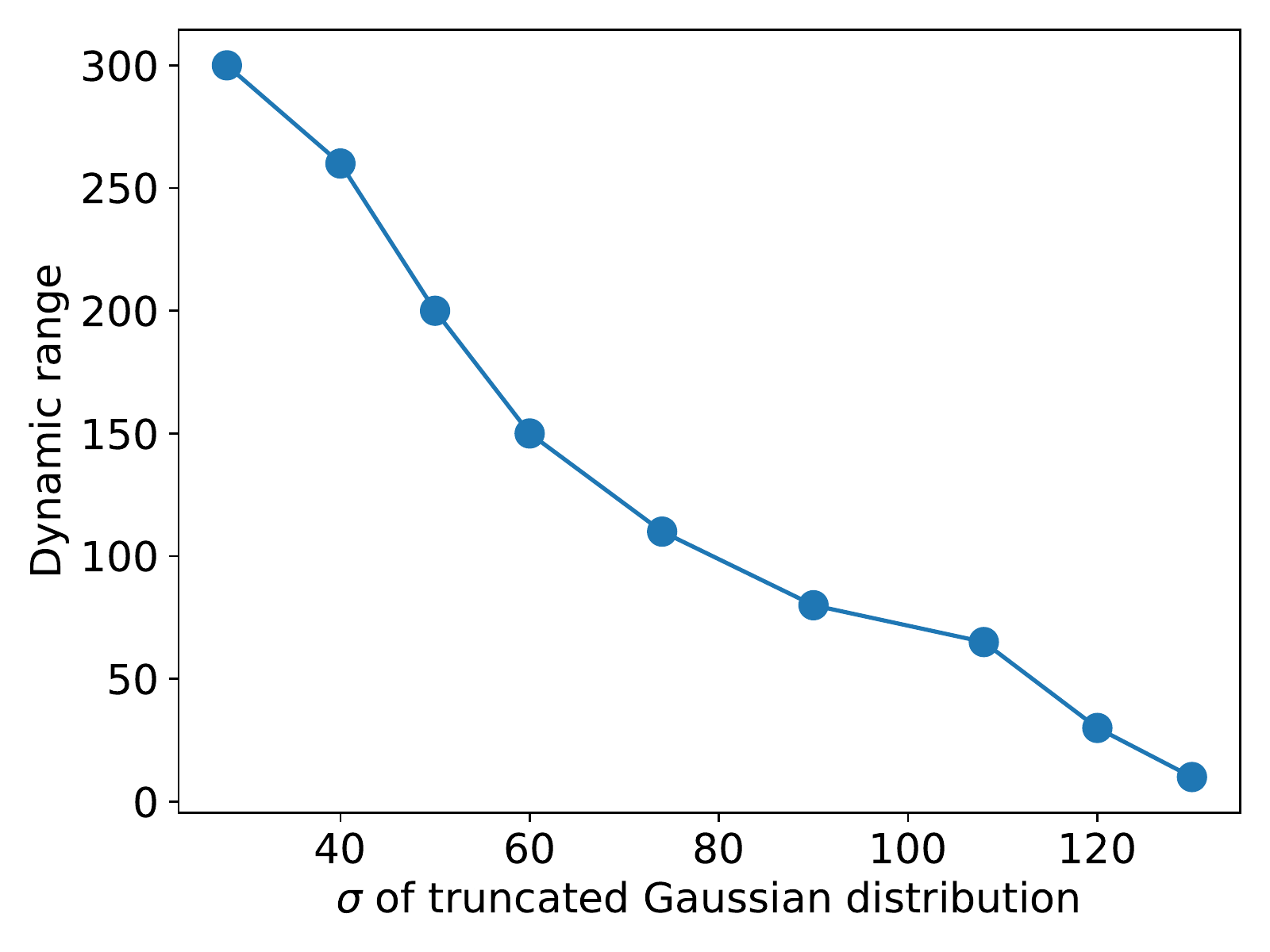}
    \caption{Variation of DR with the $\sigma$ of the truncated Gaussian distribution.}
    \label{fig:snr_with_phase}
\end{figure}

These simulations are done for a wide range of standard deviations. Here the results from three sample standard deviations of 28, 74, and 108 degrees are shown. The distribution of the simulated phases is shown in the left panels of Figure \ref{fig:gaussian}. The dirty images made from the simulated visibilities are shown in the right panels Figure \ref{fig:gaussian}. In all three situations, there is a source emission detected with more than 10-$\sigma$ detection near the phase center. The DR of the images decreases with the increase in the standard deviation of the truncated Gaussian distribution. The DR of the images are 300, 110, and 55, respectively, for the truncated Gaussian distributions with 28, 74, and 108 $\mathrm{degrees}$ standard deviation. DR is plotted against $\sigma$ in Figure \ref{fig:snr_with_phase}, which monotonically decreases with the increase in $\sigma$. It has been found that for $\sigma\geq\ 120\ \mathrm{degrees}$, the DR becomes lower than 20.

\section{Discussion and Summary}\label{sec:discussion}
AIRCARS is a self-calibration-based algorithm. Any self-calibration-based algorithm has some intrinsic limitations such as the loss of absolute flux density scale and astrometric accuracy. Both the flux density calibration and astrometric accuracy are important for cross-comparisons with observations at other wavelengths. In the past, the solar flux density calibration was done using a instrumental gain-independent method described by \citet{oberoi2017} and \citet{Mohan2017spreads}. Recently a new technique has been developed that utilizes the instrumental characterization and very stable instrumental bandpass  of the MWA \citep{Kansabanik2022}. This flux density calibration method is much more general and is now the default technique for this purpose. The astrometric correction in AIRCARS is done based on an image-based approach, which provides astrometric accuracy better than the PSF size (a few arcmins). The detailed description of that method is beyond the scope of this article and will be described in a forthcoming article \citep[][in preparation]{Kansabanik2022_paircarsII}. 

One of the novel features of AIRCARS is that it can perform the calibration of solar observations with the MWA even without any dedicated calibrator observations exploiting the partial coherency of the MWA array. The partial phase stability is demonstrated using simulation for a Gaussian distribution (Figure \ref{fig:gaussian}), but in practice, the distribution of phase may not follow a true Gaussian distribution. But, the fact that the phase distribution (Figure \ref{fig:gain_stats}) has a strong peak, and the baselines comprising the antennas whose phase lies close to the strong peak will have always some coherency between them and hence can be used to produce a reasonably accurate source model to start the self-calibration procedure even in the absence of calibrator observation.

During solar maxima, ionospheric activities are expected to be larger. Hence it is instructive to test AIRCARS on datasets from both solar maxima and solar minima. All the examples shown in this article are from 2014 and 2015, which is close to the maxima of the Solar Cycle 24. AIRCARS has been tested on datasets covering both from solar maxima and solar minima, and has led to a large number of new discoveries \citep[e.g.][etc.]{Mohan2019a,Mohan2019b,Mondal2020b,Mondal2021a,Mondal2021b,Mohan2021a,Mohan2021b,Kansabanik2022}. This demonstrates the robustness of the AIRCARS algorithm, which is independent of the solar and ionopsheric conditions.

This article demonstrates this statistically and quantitatively. It is anticipated that AIRCARS will serve the purpose of calibration and imaging for the future radio interferometers if certain conditions are satisfied by the array:
\begin{enumerate}[label=\roman*)]
   \item Instrumental gains of all antennas should be similar. This demands precision in manufacturing of the antenna elements.
    \item The delays introduced by the electrical cables need to be measured properly at regular intervals and should be corrected before performing the cross-correlation. This will reduce the loss of coherency.
    \item A large number of antennas need to be distributed over a small array footprint, such that all the baselines originated from the core are dominated by the core$-$core baselines independent of the array footprint.
\end{enumerate}

Among these three criteria, the third criterion depends on the array configuration. This is expected to be satisfied by the future SKA and some other next-generation radio interferometers; like the \textit{Next Generation Very Large Array} \citep[ngVLA:][]{ngVLA2019}, and the \textit{Frequency Agile Solar Radiotelescope} \citep[FASR:][]{Gary2003,Bastian2005,Bastian2019}.
ngVLA is planned to observe at $1-115\ \mathrm{GHz}$ and has three separate array configurations that operate in parallel. Among these three array configurations, the \textit{Short Baseline Array} (SBA) consists of $19\times6\ \mathrm{m}$ antennas located at the current VLA site and is highly suited for high-fidelity spectroscopic snapshot imaging of the Sun. FASR will be a solar dedicated radio interferometer operating at $0.2-20\ \mathrm{GHz}$. Two separate array configurations have been proposed \citep{Bastian2019} for FASR, which provide dense {\it uv-}coverage over the large bandwidth. The array footprint of FASR is similar to the MWA Phase-I, hence AIRCARS is expected to work efficiently on future FASR observations (S. Mondal, private communication, 2022).

\section*{Acknowledgments}
This work makes use of the \textit{Murchison Radio-astronomy Observatory} (MRO), operated by the Commonwealth Scientific and Industrial Research Organisation (CSIRO). D. Kansabanik acknowledges the Wajarri Yamatji people as the traditional owners of the Observatory site. Support for the operation of the MWA is provided by the Australian Government's National Collaborative Research Infrastructure Strategy (NCRIS), under a contract to Curtin University administered by Astronomy Australia Limited. D. Kansabanik acknowledges the Pawsey Supercomputing Centre, which is supported by the Western Australian and Australian Governments. D. Kansabanik acknowledges the support of the Department of Atomic Energy, Government of India, under project no. 12-R\&D-TFR-5.02-0700. D. Kansabanik gratefully acknowledges Divya Oberoi (NCRA-TIFR) and Surajit Mondal (NJIT, USA) for useful discussion, suggestions, and help in improving the manuscript. D. Kansabanik thanks the anonymous reviewer for helpful comments, which have helped to improve the clarity and the presentation of this work. 

\section*{Data Availability}
MWA solar observations used in this work came from 05 May 2014, and 11 November 2015 under the project id G0002, which are publicly available at \textsf{asvo.mwatelescope.org}.

\section*{Conflicts of Interest}
The author declares that there are no commercial or financial
relationships that could be regarded as a potential conflict of interest.



\bibliographystyle{spr-mp-sola}
\bibliography{manuscript} 

\begin{thebibliography}{43}
\ifx\bisbn     \undefined \def\bisbn  #1{ISBN #1}\fi
\ifx\binits    \undefined \def\binits#1{#1}\fi
\ifx\bauthor   \undefined \def\bauthor#1{#1}\fi
\ifx\batitle   \undefined \def\batitle#1{#1}\fi
\ifx\bjtitle   \undefined \def\bjtitle#1{\textit{#1}}\fi
\ifx\bvolume   \undefined \def\bvolume#1{\textbf{#1}}\fi
\ifx\byear     \undefined \def\byear#1{#1}\fi
\ifx\bissue    \undefined \def\bissue#1{#1}\fi
\ifx\bfpage    \undefined \def\bfpage#1{#1}\fi
\ifx\blpage    \undefined \def\blpage #1{#1}\fi
\ifx\burl      \undefined \def\burl#1{#1}\fi
\ifx\href      \undefined \def\href#1#2{#2}\fi
\ifx\betal     \undefined \def\betal{et al.}\fi
\ifx\bctitle   \undefined \def\bctitle#1{#1}\fi
\ifx\beditor   \undefined \def\beditor#1{#1}\fi
\ifx\bbtitle   \undefined \def\bbtitle#1{\textit{#1}}\fi
\ifx\bedition  \undefined \def\bedition#1{#1}\fi
\ifx\bseriesno \undefined \def\bseriesno#1{\textbf{#1}}\fi
\ifx\blocation \undefined \def\blocation#1{#1}\fi
\ifx\bsertitle \undefined \def\bsertitle#1{\textit{#1}}\fi
\ifx\bsnm      \undefined \def\bsnm#1{#1}\fi
\ifx\bsuffix   \undefined \def\bsuffix#1{#1}\fi
\ifx\bparticle \undefined \def\bparticle#1{#1}\fi
\ifx\barticle  \undefined \def\barticle#1{}\fi
\ifx\binstitute  \undefined \def\binstitute#1{#1}\fi
\ifx\bpublisher  \undefined \def\bpublisher#1{#1}\fi
\ifx\doiurl    \undefined \def\doiurl#1{\href{#1}{DOI}}\fi
\makeatletter
\def\safeHref#1#2#3{\in@{http}{#2}\ifin@\href{#2}{#3}\else\href{#1#2}{#3}\fi}
\makeatother
\ifx\adsurl    \undefined
  \def\adsurl#1{\safeHref{https://ui.adsabs.harvard.edu/abs/}{#1}{ADS}}\fi
\ifx\arxivurl  \undefined
  \def\arxivurl#1{\safeHref{http://arxiv.org/abs/}{#1}{arXiv}}\fi
\ifx\botherref \undefined \def\botherref#1{}\fi
\ifx\url       \undefined \def\url#1{#1}\fi
\ifx\bchapter  \undefined \def\bchapter#1{}\fi
\ifx\bbook     \undefined \def\bbook#1{}\fi
\ifx\bcomment  \undefined \def\bcomment#1{#1}\fi
\ifx\oauthor   \undefined \def\oauthor#1{#1}\fi
\ifx\citeauthoryear \undefined\def \citeauthoryear#1{#1}\fi
\def\endbibitem {}
\ifx\bconflocation  \undefined \def\bconflocation#1{#1} \fi

\bibitem[\protect\citeauthoryear{Bastian}{2005}]{Bastian2005}
\begin{bchapter}
\bauthor{\bsnm{Bastian}, \binits{T.S.}}:
\byear{2005},
\bctitle{The Frequency Agile Solar Radiotelescope}.
In: \beditor{\bsnm{Gary}, \binits{D.E.}},
\beditor{\bsnm{Keller}, \binits{C.U.}} (eds.)
\bbtitle{Solar and Space Weather Radiophysics},
\bpublisher{Astrophy. Space Sci. Lib. 314, Springer Netherlands},
\blocation{Dordrecht},
\bfpage{47}.
\doiurl{https://doi.org/10.1007/1-4020-2814-8_3}.
\end{bchapter}
\endbibitem

\bibitem[\protect\citeauthoryear{{Bastian} et~al.}{2001}]{bastian2001}
\begin{barticle}
\bauthor{\bsnm{{Bastian}}, \binits{T.S.}},
\bauthor{\bsnm{{Pick}}, \binits{M.}},
\bauthor{\bsnm{{Kerdraon}}, \binits{A.}},
\bauthor{\bsnm{{Maia}}, \binits{D.}},
\bauthor{\bsnm{{Vourlidas}}, \binits{A.}}:
\byear{2001},
\batitle{{The Coronal Mass Ejection of 1998 April 20: Direct Imaging at Radio
  Wavelengths}}.
\bjtitle{\apjl}
\bvolume{558},
\bfpage{L65}.
\doiurl{https://doi.org/10.1086/323421}.
\adsurl{2001ApJ...558L..65B}.
\end{barticle}
\endbibitem

\bibitem[\protect\citeauthoryear{{Bastian} et~al.}{2019}]{Bastian2019}
\begin{barticle}
\bauthor{\bsnm{{Bastian}}, \binits{T.}},
\bauthor{\bsnm{{Bain}}, \binits{H.}},
\bauthor{\bsnm{{Bradley}}, \binits{R.}},
\bauthor{\bsnm{{Chen}}, \binits{B.}},
\bauthor{\bsnm{{Dahlin}}, \binits{J.}},
\bauthor{\bsnm{{DeLuca}}, \binits{E.}},
\bauthor{\bsnm{{Drake}}, \binits{J.}},
\bauthor{\bsnm{{Fleishman}}, \binits{G.}},
\bauthor{\bsnm{{Gary}}, \binits{D.}},
\bauthor{\bsnm{{Glesener}}, \binits{L.}},
\bauthor{\bsnm{{Guo}}, \binits{F.}},
\bauthor{\bsnm{{Hallinan}}, \binits{G.}},
\bauthor{\bsnm{{Hurford}}, \binits{G.}},
\bauthor{\bsnm{{Kasper}}, \binits{J.}},
\bauthor{\bsnm{{Ji}}, \binits{H.}},
\bauthor{\bsnm{{Klimchuk}}, \binits{J.}},
\bauthor{\bsnm{{Kobelski}}, \binits{A.}},
\bauthor{\bsnm{{Krucker}}, \binits{S.}},
\bauthor{\bsnm{{Kuroda}}, \binits{N.}},
\bauthor{\bsnm{{Loncope}}, \binits{D.}},
\bauthor{\bsnm{{Lonsdale}}, \binits{C.}},
\bauthor{\bsnm{{McTiernan}}, \binits{J.}},
\bauthor{\bsnm{{Nita}}, \binits{G.}},
\bauthor{\bsnm{{Qiu}}, \binits{J.}},
\bauthor{\bsnm{{Reeves}}, \binits{K.}},
\bauthor{\bsnm{{Saint-Hilaire}}, \binits{P.}},
\bauthor{\bsnm{{Schonfeld}}, \binits{S.}},
\bauthor{\bsnm{{Shen}}, \binits{C.}},
\bauthor{\bsnm{{Tun}}, \binits{S.}},
\bauthor{\bsnm{{Wertheimer}}, \binits{D.}},
\bauthor{\bsnm{{White}}, \binits{S.}}:
\byear{2019},
\batitle{{Frequency Agile Solar Radiotelescope}}.
\bjtitle{Astro2020: Decadal Survey on Astronomy and Astrophysics}
\bvolume{2020},
\bfpage{56}.
\adsurl{2019astro2020U..56B}.
\end{barticle}
\endbibitem

\bibitem[\protect\citeauthoryear{Cornwell}{2008}]{cornwell2008}
\begin{barticle}
\bauthor{\bsnm{Cornwell}, \binits{T.J.}}:
\byear{2008},
\batitle{Multiscale CLEAN Deconvolution of Radio Synthesis Images}.
\bjtitle{IEEE J. Sel. Topics. Signal. Proc.}
\bvolume{2},
\bfpage{793}.
\doiurl{https://doi.org/10.1109/JSTSP.2008.2006388}.
\end{barticle}
\endbibitem

\bibitem[\protect\citeauthoryear{{Di Francesco} et~al.}{2019}]{ngVLA2019}
\begin{bchapter}
\bauthor{\bsnm{{Di Francesco}}, \binits{J.}},
\bauthor{\bsnm{{Chalmers}}, \binits{D.}},
\bauthor{\bsnm{{Denman}}, \binits{N.}},
\bauthor{\bsnm{{Fissel}}, \binits{L.}},
\bauthor{\bsnm{{Friesen}}, \binits{R.}},
\bauthor{\bsnm{{Gaensler}}, \binits{B.}},
\bauthor{\bsnm{{Hlavacek-Larrondo}}, \binits{J.}},
\bauthor{\bsnm{{Kirk}}, \binits{H.}},
\bauthor{\bsnm{{Matthews}}, \binits{B.}},
\bauthor{\bsnm{{O'Dea}}, \binits{C.}},
\bauthor{\bsnm{{Robishaw}}, \binits{T.}},
\bauthor{\bsnm{{Rosolowsky}}, \binits{E.}},
\bauthor{\bsnm{{Rupen}}, \binits{M.}},
\bauthor{\bsnm{{Sadavoy}}, \binits{S.}},
\bauthor{\bsnm{{Sa-Harb}}, \binits{S.}},
\bauthor{\bsnm{{Sivakoff}}, \binits{G.}},
\bauthor{\bsnm{{Tahani}}, \binits{M.}},
\bauthor{\bsnm{{van der Marel}}, \binits{N.}},
\bauthor{\bsnm{{White}}, \binits{J.}},
\bauthor{\bsnm{{Wilson}}, \binits{C.}}:
\byear{2019},
\bctitle{{The Next Generation Very Large Array}}.
In: \bbtitle{Canadian Long Range Plan for Astronomy and Astrophysics White
  Papers}
\bseriesno{2020},
\bfpage{32}.
\doiurl{https://doi.org/10.5281/zenodo.3765763}.
\adsurl{2019clrp.2020...32D}.
\end{bchapter}
\endbibitem

\bibitem[\protect\citeauthoryear{{Gary}}{2003}]{Gary2003}
\begin{barticle}
\bauthor{\bsnm{{Gary}}, \binits{D.E.}}:
\byear{2003},
\batitle{{The Frequency Agile Solar Radiotelescope}}.
\bjtitle{J. Korean Astron. Soc.}
\bvolume{36},
\bfpage{S135}.
\doiurl{https://doi.org/10.5303/JKAS.2003.36.spc1.135}.
\adsurl{2003JKAS...36S.135G}.
\end{barticle}
\endbibitem

\bibitem[\protect\citeauthoryear{Gupta et~al.}{2017}]{Gupta_2017}
\begin{barticle}
\bauthor{\bsnm{Gupta}, \binits{Y.}},
\bauthor{\bsnm{Ajithkumar}, \binits{B.}},
\bauthor{\bsnm{Kale}, \binits{H.S.}},
\bauthor{\bsnm{Nayak}, \binits{S.}},
\bauthor{\bsnm{Sabhapathy}, \binits{S.}},
\bauthor{\bsnm{S.}, \binits{S.}},
\bauthor{\bsnm{Swami}, \binits{R.V.}},
\bauthor{\bsnm{Chengalur}, \binits{J.}},
\bauthor{\bsnm{Ghosh}, \binits{S.K.}},
\bauthor{\bsnm{Ishwara-Chandra}, \binits{C.H.}},
\bauthor{\bsnm{Joshi}, \binits{B.}},
\bauthor{\bsnm{Kanekar}, \binits{N.}},
\bauthor{\bsnm{Lal}, \binits{D.}},
\bauthor{\bsnm{Roy}, \binits{S.}}:
\byear{2017},
\batitle{The upgraded {GMRT}: Opening new windows on the radio Universe}.
\bjtitle{Curr. Sci.}
\bvolume{113},
\bfpage{707}.
\doiurl{https://doi.org/10.18520/cs/v113/i04/707-714}.
\end{barticle}
\endbibitem

\bibitem[\protect\citeauthoryear{{Hamaker}, {Bregman}, and
  {Sault}}{1996}]{Hamaker1996_1}
\begin{barticle}
\bauthor{\bsnm{{Hamaker}}, \binits{J.P.}},
\bauthor{\bsnm{{Bregman}}, \binits{J.D.}},
\bauthor{\bsnm{{Sault}}, \binits{R.J.}}:
\byear{1996},
\batitle{{Understanding radio polarimetry. I. Mathematical foundations.}}
\bjtitle{\aaps}
\bvolume{117},
\bfpage{137}.
\adsurl{1996A&AS..117..137H}.
\end{barticle}
\endbibitem

\bibitem[\protect\citeauthoryear{{Hurley-Walker} and
  {Hancock}}{2018}]{Walker2018}
\begin{barticle}
\bauthor{\bsnm{{Hurley-Walker}}, \binits{N.}},
\bauthor{\bsnm{{Hancock}}, \binits{P.J.}}:
\byear{2018},
\batitle{{De-distorting ionospheric effects in the image plane}}.
\bjtitle{Astron. Compu.}
\bvolume{25},
\bfpage{94}.
\doiurl{https://doi.org/10.1016/j.ascom.2018.08.006}.
\adsurl{2018A&C....25...94H}.
\end{barticle}
\endbibitem

\bibitem[\protect\citeauthoryear{Kansabanik, Oberoi, and
  Mondal}{2022a}]{Kansabanik2022_GS}
\begin{botherref}
\oauthor{\bsnm{Kansabanik}, \binits{D.}},
\oauthor{\bsnm{Oberoi}, \binits{D.}},
\oauthor{\bsnm{Mondal}, \binits{S.}}:
2022a,
Deciphering the Faint Gyrosynchrotron Emission from Coronal Mass Ejections with
  Low-frequency Spectro-polarimetric Radio Observation.
\textit{\apj}.
\end{botherref}
\endbibitem

\bibitem[\protect\citeauthoryear{Kansabanik, Oberoi, and
  Mondal}{2022b}]{Kansabanik2022_paircarsII}
\begin{botherref}
\oauthor{\bsnm{Kansabanik}, \binits{D.}},
\oauthor{\bsnm{Oberoi}, \binits{D.}},
\oauthor{\bsnm{Mondal}, \binits{S.}}:
2022b,
Tackling the Unique Challenges of Low-frequency Solar Polarimetry with the
  Square Kilometre Array Low Precursor: Pipeline Implemntation.
\textit{\apj}.
\end{botherref}
\endbibitem

\bibitem[\protect\citeauthoryear{Kansabanik, Oberoi, and
  Mondal}{2022c}]{Kansabanik2022_paircarsI}
\begin{barticle}
\bauthor{\bsnm{Kansabanik}, \binits{D.}},
\bauthor{\bsnm{Oberoi}, \binits{D.}},
\bauthor{\bsnm{Mondal}, \binits{S.}}:
\byear{2022}c,
\batitle{Tackling the Unique Challenges of Low-frequency Solar Polarimetry with
  the Square Kilometre Array Low Precursor: The Algorithm}.
\bjtitle{\apj}
\bvolume{932},
\bfpage{110}.
\doiurl{https://doi.org/10.3847/1538-4357/ac6758}.
\end{barticle}
\endbibitem

\bibitem[\protect\citeauthoryear{Kansabanik et~al.}{2022}]{Kansabanik2022}
\begin{barticle}
\bauthor{\bsnm{Kansabanik}, \binits{D.}},
\bauthor{\bsnm{Mondal}, \binits{S.}},
\bauthor{\bsnm{Oberoi}, \binits{D.}},
\bauthor{\bsnm{Biswas}, \binits{A.}},
\bauthor{\bsnm{Bhunia}, \binits{S.}}:
\byear{2022},
\batitle{Robust Absolute Solar Flux Density Calibration for the Murchison
  Widefield Array}.
\bjtitle{\apj}
\bvolume{927},
\bfpage{17}.
\doiurl{https://doi.org/10.3847/1538-4357/ac4bba}.
\end{barticle}
\endbibitem

\bibitem[\protect\citeauthoryear{{Line} et~al.}{2018}]{Line2018}
\begin{barticle}
\bauthor{\bsnm{{Line}}, \binits{J.L.B.}},
\bauthor{\bsnm{{McKinley}}, \binits{B.}},
\bauthor{\bsnm{{Rasti}}, \binits{J.}},
\bauthor{\bsnm{{Bhardwaj}}, \binits{M.}},
\bauthor{\bsnm{{Wayth}}, \binits{R.B.}},
\bauthor{\bsnm{{Webster}}, \binits{R.L.}},
\bauthor{\bsnm{{Ung}}, \binits{D.}},
\bauthor{\bsnm{{Emrich}}, \binits{D.}},
\bauthor{\bsnm{{Horsley}}, \binits{L.}},
\bauthor{\bsnm{{Beardsley}}, \binits{A.}},
\bauthor{\bsnm{{Crosse}}, \binits{B.}},
\bauthor{\bsnm{{Franzen}}, \binits{T.M.O.}},
\bauthor{\bsnm{{Gaensler}}, \binits{B.M.}},
\bauthor{\bsnm{{Johnston-Hollitt}}, \binits{M.}},
\bauthor{\bsnm{{Kaplan}}, \binits{D.L.}},
\bauthor{\bsnm{{Kenney}}, \binits{D.}},
\bauthor{\bsnm{{Morales}}, \binits{M.F.}},
\bauthor{\bsnm{{Pallot}}, \binits{D.}},
\bauthor{\bsnm{{Steele}}, \binits{K.}},
\bauthor{\bsnm{{Tingay}}, \binits{S.J.}},
\bauthor{\bsnm{{Trott}}, \binits{C.M.}},
\bauthor{\bsnm{{Walker}}, \binits{M.}},
\bauthor{\bsnm{{Williams}}, \binits{A.}},
\bauthor{\bsnm{{Wu}}, \binits{C.}}:
\byear{2018},
\batitle{{In situ measurement of MWA primary beam variation using ORBCOMM}}.
\bjtitle{\pasa}
\bvolume{35},
\bfpage{e045}.
\doiurl{https://doi.org/10.1017/pasa.2018.30}.
\adsurl{2018PASA...35...45L}.
\end{barticle}
\endbibitem

\bibitem[\protect\citeauthoryear{{Loi} et~al.}{2015a}]{Loi2015a}
\begin{barticle}
\bauthor{\bsnm{{Loi}}, \binits{S.T.}},
\bauthor{\bsnm{{Trott}}, \binits{C.M.}},
\bauthor{\bsnm{{Murphy}}, \binits{T.}},
\bauthor{\bsnm{{Cairns}}, \binits{I.H.}},
\bauthor{\bsnm{{Bell}}, \binits{M.}},
\bauthor{\bsnm{{Hurley-Walker}}, \binits{N.}},
\bauthor{\bsnm{{Morgan}}, \binits{J.}},
\bauthor{\bsnm{{Lenc}}, \binits{E.}},
\bauthor{\bsnm{{Offringa}}, \binits{A.R.}},
\bauthor{\bsnm{{Feng}}, \binits{L.}},
\bauthor{\bsnm{{Hancock}}, \binits{P.J.}},
\bauthor{\bsnm{{Kaplan}}, \binits{D.L.}},
\bauthor{\bsnm{{Kudryavtseva}}, \binits{N.}},
\bauthor{\bsnm{{Bernardi}}, \binits{G.}},
\bauthor{\bsnm{{Bowman}}, \binits{J.D.}},
\bauthor{\bsnm{{Briggs}}, \binits{F.}},
\bauthor{\bsnm{{Cappallo}}, \binits{R.J.}},
\bauthor{\bsnm{{Corey}}, \binits{B.E.}},
\bauthor{\bsnm{{Deshpande}}, \binits{A.A.}},
\bauthor{\bsnm{{Emrich}}, \binits{D.}},
\bauthor{\bsnm{{Gaensler}}, \binits{B.M.}},
\bauthor{\bsnm{{Goeke}}, \binits{R.}},
\bauthor{\bsnm{{Greenhill}}, \binits{L.J.}},
\bauthor{\bsnm{{Hazelton}}, \binits{B.J.}},
\bauthor{\bsnm{{Johnston-Hollitt}}, \binits{M.}},
\bauthor{\bsnm{{Kasper}}, \binits{J.C.}},
\bauthor{\bsnm{{Kratzenberg}}, \binits{E.}},
\bauthor{\bsnm{{Lonsdale}}, \binits{C.J.}},
\bauthor{\bsnm{{Lynch}}, \binits{M.J.}},
\bauthor{\bsnm{{McWhirter}}, \binits{S.R.}},
\bauthor{\bsnm{{Mitchell}}, \binits{D.A.}},
\bauthor{\bsnm{{Morales}}, \binits{M.F.}},
\bauthor{\bsnm{{Morgan}}, \binits{E.}},
\bauthor{\bsnm{{Oberoi}}, \binits{D.}},
\bauthor{\bsnm{{Ord}}, \binits{S.M.}},
\bauthor{\bsnm{{Prabu}}, \binits{T.}},
\bauthor{\bsnm{{Rogers}}, \binits{A.E.E.}},
\bauthor{\bsnm{{Roshi}}, \binits{A.}},
\bauthor{\bsnm{{Shankar}}, \binits{N.U.}},
\bauthor{\bsnm{{Srivani}}, \binits{K.S.}},
\bauthor{\bsnm{{Subrahmanyan}}, \binits{R.}},
\bauthor{\bsnm{{Tingay}}, \binits{S.J.}},
\bauthor{\bsnm{{Waterson}}, \binits{M.}},
\bauthor{\bsnm{{Wayth}}, \binits{R.B.}},
\bauthor{\bsnm{{Webster}}, \binits{R.L.}},
\bauthor{\bsnm{{Whitney}}, \binits{A.R.}},
\bauthor{\bsnm{{Williams}}, \binits{A.}},
\bauthor{\bsnm{{Williams}}, \binits{C.L.}}:
\byear{2015}a,
\batitle{{Power spectrum analysis of ionospheric fluctuations with the
  Murchison Widefield Array}}.
\bjtitle{Radio Sci.}
\bvolume{50},
\bfpage{574}.
\doiurl{https://doi.org/10.1002/2015RS005711}.
\adsurl{2015RaSc...50..574L}.
\end{barticle}
\endbibitem

\bibitem[\protect\citeauthoryear{{Loi} et~al.}{2015b}]{Loi2015b}
\begin{barticle}
\bauthor{\bsnm{{Loi}}, \binits{S.T.}},
\bauthor{\bsnm{{Murphy}}, \binits{T.}},
\bauthor{\bsnm{{Cairns}}, \binits{I.H.}},
\bauthor{\bsnm{{Menk}}, \binits{F.W.}},
\bauthor{\bsnm{{Waters}}, \binits{C.L.}},
\bauthor{\bsnm{{Erickson}}, \binits{P.J.}},
\bauthor{\bsnm{{Trott}}, \binits{C.M.}},
\bauthor{\bsnm{{Hurley-Walker}}, \binits{N.}},
\bauthor{\bsnm{{Morgan}}, \binits{J.}},
\bauthor{\bsnm{{Lenc}}, \binits{E.}},
\bauthor{\bsnm{{Offringa}}, \binits{A.R.}},
\bauthor{\bsnm{{Bell}}, \binits{M.E.}},
\bauthor{\bsnm{{Ekers}}, \binits{R.D.}},
\bauthor{\bsnm{{Gaensler}}, \binits{B.M.}},
\bauthor{\bsnm{{Lonsdale}}, \binits{C.J.}},
\bauthor{\bsnm{{Feng}}, \binits{L.}},
\bauthor{\bsnm{{Hancock}}, \binits{P.J.}},
\bauthor{\bsnm{{Kaplan}}, \binits{D.L.}},
\bauthor{\bsnm{{Bernardi}}, \binits{G.}},
\bauthor{\bsnm{{Bowman}}, \binits{J.D.}},
\bauthor{\bsnm{{Briggs}}, \binits{F.}},
\bauthor{\bsnm{{Cappallo}}, \binits{R.J.}},
\bauthor{\bsnm{{Deshpande}}, \binits{A.A.}},
\bauthor{\bsnm{{Greenhill}}, \binits{L.J.}},
\bauthor{\bsnm{{Hazelton}}, \binits{B.J.}},
\bauthor{\bsnm{{Johnston-Hollitt}}, \binits{M.}},
\bauthor{\bsnm{{McWhirter}}, \binits{S.R.}},
\bauthor{\bsnm{{Mitchell}}, \binits{D.A.}},
\bauthor{\bsnm{{Morales}}, \binits{M.F.}},
\bauthor{\bsnm{{Morgan}}, \binits{E.}},
\bauthor{\bsnm{{Oberoi}}, \binits{D.}},
\bauthor{\bsnm{{Ord}}, \binits{S.M.}},
\bauthor{\bsnm{{Prabu}}, \binits{T.}},
\bauthor{\bsnm{{Shankar}}, \binits{N.U.}},
\bauthor{\bsnm{{Srivani}}, \binits{K.S.}},
\bauthor{\bsnm{{Subrahmanyan}}, \binits{R.}},
\bauthor{\bsnm{{Tingay}}, \binits{S.J.}},
\bauthor{\bsnm{{Wayth}}, \binits{R.B.}},
\bauthor{\bsnm{{Webster}}, \binits{R.L.}},
\bauthor{\bsnm{{Williams}}, \binits{A.}},
\bauthor{\bsnm{{Williams}}, \binits{C.L.}}:
\byear{2015}b,
\batitle{{Real-time imaging of density ducts between the plasmasphere and
  ionosphere}}.
\bjtitle{\grl}
\bvolume{42},
\bfpage{3707}.
\doiurl{https://doi.org/10.1002/2015GL063699}.
\adsurl{2015GeoRL..42.3707L}.
\end{barticle}
\endbibitem

\bibitem[\protect\citeauthoryear{{Lonsdale}}{2005}]{Lonsdale2005}
\begin{bchapter}
\bauthor{\bsnm{{Lonsdale}}, \binits{C.J.}}:
\byear{2005},
\bctitle{{Configuration Considerations for Low Frequency Arrays}}.
In: \beditor{\bsnm{{Kassim}}, \binits{N.}},
\beditor{\bsnm{{Perez}}, \binits{M.}},
\beditor{\bsnm{{Junor}}, \binits{W.}},
\beditor{\bsnm{{Henning}}, \binits{P.}} (eds.)
\bbtitle{From Clark Lake to the Long Wavelength Array: Bill Erickson's Radio
  Sci.},
\bsertitle{Astron. Soc. Pacific. CS.}
\bseriesno{345},
\bconflocation{San Francisco},
\bfpage{399}.
\adsurl{2005ASPC..345..399L/abstract}.
\end{bchapter}
\endbibitem

\bibitem[\protect\citeauthoryear{{McCready}, {Pawsey}, and
  {Payne-Scott}}{1947}]{McCready1947}
\begin{barticle}
\bauthor{\bsnm{{McCready}}, \binits{L.L.}},
\bauthor{\bsnm{{Pawsey}}, \binits{J.L.}},
\bauthor{\bsnm{{Payne-Scott}}, \binits{R.}}:
\byear{1947},
\batitle{{Solar Radiation at Radio Frequencies and Its Relation to Sunspots}}.
\bjtitle{Proc. Royal Soc. London Ser. A}
\bvolume{190},
\bfpage{357}.
\doiurl{https://doi.org/10.1098/rspa.1947.0081}.
\adsurl{1947RSPSA.190..357M}.
\end{barticle}
\endbibitem

\bibitem[\protect\citeauthoryear{{McLean} and {Labrum}}{1985}]{McLeanBook}
\begin{bbook}
\bauthor{\bsnm{{McLean}}, \binits{D.J.}},
\bauthor{\bsnm{{Labrum}}, \binits{N.R.}}:
\byear{1985},
\bbtitle{Solar radiophysics : studies of emission from the sun at metre
  wavelengths},
\bpublisher{Cambridge University Press},
\blocation{Cambridge}.
\adsurl{1985srph.book.....M}.
\end{bbook}
\endbibitem

\bibitem[\protect\citeauthoryear{{McMullin} et~al.}{2007}]{mcmullin2007}
\begin{bchapter}
\bauthor{\bsnm{{McMullin}}, \binits{J.P.}},
\bauthor{\bsnm{{Waters}}, \binits{B.}},
\bauthor{\bsnm{{Schiebel}}, \binits{D.}},
\bauthor{\bsnm{{Young}}, \binits{W.}},
\bauthor{\bsnm{{Golap}}, \binits{K.}}:
\byear{2007},
\bctitle{{CASA Architecture and Applications}}.
In: \beditor{\bsnm{{Shaw}}, \binits{R.A.}},
\beditor{\bsnm{{Hill}}, \binits{F.}},
\beditor{\bsnm{{Bell}}, \binits{D.J.}} (eds.)
\bbtitle{Astronomical Data Analysis Software and Systems XVI},
\bsertitle{Astron. Soc. Pacific. CS.}
\bseriesno{376},
\bconflocation{San Francisco},
\bfpage{127}.
\adsurl{2007ASPC..376..127M}.
\end{bchapter}
\endbibitem

\bibitem[\protect\citeauthoryear{{Mevius} et~al.}{2016}]{Mevius2015}
\begin{barticle}
\bauthor{\bsnm{{Mevius}}, \binits{M.}},
\bauthor{\bsnm{{van der Tol}}, \binits{S.}},
\bauthor{\bsnm{{Pandey}}, \binits{V.N.}},
\bauthor{\bsnm{{Vedantham}}, \binits{H.K.}},
\bauthor{\bsnm{{Brentjens}}, \binits{M.A.}},
\bauthor{\bsnm{{de Bruyn}}, \binits{A.G.}},
\bauthor{\bsnm{{Abdalla}}, \binits{F.B.}},
\bauthor{\bsnm{{Asad}}, \binits{K.M.B.}},
\bauthor{\bsnm{{Bregman}}, \binits{J.D.}},
\bauthor{\bsnm{{Brouw}}, \binits{W.N.}},
\bauthor{\bsnm{{Bus}}, \binits{S.}},
\bauthor{\bsnm{{Chapman}}, \binits{E.}},
\bauthor{\bsnm{{Ciardi}}, \binits{B.}},
\bauthor{\bsnm{{Fernandez}}, \binits{E.R.}},
\bauthor{\bsnm{{Ghosh}}, \binits{A.}},
\bauthor{\bsnm{{Harker}}, \binits{G.}},
\bauthor{\bsnm{{Iliev}}, \binits{I.T.}},
\bauthor{\bsnm{{Jeli{\'c}}}, \binits{V.}},
\bauthor{\bsnm{{Kazemi}}, \binits{S.}},
\bauthor{\bsnm{{Koopmans}}, \binits{L.V.E.}},
\bauthor{\bsnm{{Noordam}}, \binits{J.E.}},
\bauthor{\bsnm{{Offringa}}, \binits{A.R.}},
\bauthor{\bsnm{{Patil}}, \binits{A.H.}},
\bauthor{\bsnm{{van Weeren}}, \binits{R.J.}},
\bauthor{\bsnm{{Wijnholds}}, \binits{S.}},
\bauthor{\bsnm{{Yatawatta}}, \binits{S.}},
\bauthor{\bsnm{{Zaroubi}}, \binits{S.}}:
\byear{2016},
\batitle{{Probing ionospheric structures using the LOFAR radio telescope}}.
\bjtitle{Radio Sci.}
\bvolume{51},
\bfpage{927}.
\doiurl{https://doi.org/10.1002/2016RS006028}.
\adsurl{2016RaSc...51..927M}.
\end{barticle}
\endbibitem

\bibitem[\protect\citeauthoryear{Mitchell et~al.}{2008}]{Mitchell2008}
\begin{barticle}
\bauthor{\bsnm{Mitchell}, \binits{D.A.}},
\bauthor{\bsnm{Greenhill}, \binits{L.J.}},
\bauthor{\bsnm{Wayth}, \binits{R.B.}},
\bauthor{\bsnm{Sault}, \binits{R.J.}},
\bauthor{\bsnm{Lonsdale}, \binits{C.J.}},
\bauthor{\bsnm{Cappallo}, \binits{R.J.}},
\bauthor{\bsnm{Morales}, \binits{M.F.}},
\bauthor{\bsnm{Ord}, \binits{S.M.}}:
\byear{2008},
\batitle{Real-Time Calibration of the Murchison Widefield Array}.
\bjtitle{IEEE J. Sel. Topics. Signal Proc.}
\bvolume{2},
\bfpage{707}.
\doiurl{https://doi.org/10.1109/JSTSP.2008.2005327}.
\end{barticle}
\endbibitem

\bibitem[\protect\citeauthoryear{{Mohan}}{2021a}]{Mohan2021b}
\begin{barticle}
\bauthor{\bsnm{{Mohan}}, \binits{A.}}:
\byear{2021}a,
\batitle{{Characterising coronal turbulence using snapshot imaging of radio
  bursts in 80-200 MHz}}.
\bjtitle{\aap}
\bvolume{655},
\bfpage{A77}.
\doiurl{https://doi.org/10.1051/0004-6361/202142029}.
\adsurl{2021A&A...655A..77M}.
\end{barticle}
\endbibitem

\bibitem[\protect\citeauthoryear{{Mohan}}{2021b}]{Mohan2021a}
\begin{barticle}
\bauthor{\bsnm{{Mohan}}, \binits{A.}}:
\byear{2021}b,
\batitle{{Discovery of Correlated Evolution in Solar Noise Storm Source
  Parameters: Insights on Magnetic Field Dynamics during a Microflare}}.
\bjtitle{\apjl}
\bvolume{909},
\bfpage{L1}.
\doiurl{https://doi.org/10.3847/2041-8213/abe70a}.
\adsurl{2021ApJ...909L...1M}.
\end{barticle}
\endbibitem

\bibitem[\protect\citeauthoryear{{Mohan} and {Oberoi}}{2017}]{Mohan2017spreads}
\begin{barticle}
\bauthor{\bsnm{{Mohan}}, \binits{A.}},
\bauthor{\bsnm{{Oberoi}}, \binits{D.}}:
\byear{2017},
\batitle{{4D Data Cubes from Radio-Interferometric Spectroscopic Snapshot
  Imaging}}.
\bjtitle{\solphys}
\bvolume{292},
\bfpage{168}.
\doiurl{https://doi.org/10.1007/s11207-017-1193-1}.
\adsurl{2017SoPh..292..168M}.
\end{barticle}
\endbibitem

\bibitem[\protect\citeauthoryear{{Mohan} et~al.}{2019a}]{Mohan2019b}
\begin{barticle}
\bauthor{\bsnm{{Mohan}}, \binits{A.}},
\bauthor{\bsnm{{McCauley}}, \binits{P.I.}},
\bauthor{\bsnm{{Oberoi}}, \binits{D.}},
\bauthor{\bsnm{{Mastrano}}, \binits{A.}}:
\byear{2019}a,
\batitle{{A Weak Coronal Heating Event Associated with Periodic Particle
  Acceleration Episodes}}.
\bjtitle{\apj}
\bvolume{883},
\bfpage{45}.
\doiurl{https://doi.org/10.3847/1538-4357/ab3a94}.
\adsurl{2019ApJ...883...45M}.
\end{barticle}
\endbibitem

\bibitem[\protect\citeauthoryear{{Mohan} et~al.}{2019b}]{Mohan2019a}
\begin{barticle}
\bauthor{\bsnm{{Mohan}}, \binits{A.}},
\bauthor{\bsnm{{Mondal}}, \binits{S.}},
\bauthor{\bsnm{{Oberoi}}, \binits{D.}},
\bauthor{\bsnm{{Lonsdale}}, \binits{C.J.}}:
\byear{2019}b,
\batitle{{Evidence for Super-Alfv{\'e}nic Oscillations in Solar Type III Radio
  Burst Sources}}.
\bjtitle{\apj}
\bvolume{875},
\bfpage{98}.
\doiurl{https://doi.org/10.3847/1538-4357/ab0ae5}.
\adsurl{2019ApJ...875...98M}.
\end{barticle}
\endbibitem

\bibitem[\protect\citeauthoryear{{Mondal}}{2021}]{Mondal2021b}
\begin{barticle}
\bauthor{\bsnm{{Mondal}}, \binits{S.}}:
\byear{2021},
\batitle{{A Search for the Counterparts of Quiet-Sun Radio Transients in
  Extreme Ultraviolet Data}}.
\bjtitle{\solphys}
\bvolume{296},
\bfpage{131}.
\doiurl{https://doi.org/10.1007/s11207-021-01877-3}.
\adsurl{2021SoPh..296..131M}.
\end{barticle}
\endbibitem

\bibitem[\protect\citeauthoryear{{Mondal} and {Oberoi}}{2021}]{Mondal2021a}
\begin{barticle}
\bauthor{\bsnm{{Mondal}}, \binits{S.}},
\bauthor{\bsnm{{Oberoi}}, \binits{D.}}:
\byear{2021},
\batitle{{Insights from Snapshot Spectroscopic Radio Observations of a Weak
  Type I Solar Noise Storm}}.
\bjtitle{\apj}
\bvolume{920},
\bfpage{11}.
\doiurl{https://doi.org/10.3847/1538-4357/ac1076}.
\adsurl{2021ApJ...920...11M}.
\end{barticle}
\endbibitem

\bibitem[\protect\citeauthoryear{{Mondal}, {Oberoi}, and
  {Mohan}}{2020}]{Mondal2020b}
\begin{barticle}
\bauthor{\bsnm{{Mondal}}, \binits{S.}},
\bauthor{\bsnm{{Oberoi}}, \binits{D.}},
\bauthor{\bsnm{{Mohan}}, \binits{A.}}:
\byear{2020},
\batitle{{First Radio Evidence for Impulsive Heating Contribution to the Quiet
  Solar Corona}}.
\bjtitle{\apjl}
\bvolume{895},
\bfpage{L39}.
\doiurl{https://doi.org/10.3847/2041-8213/ab8817}.
\adsurl{2020ApJ...895L..39M}.
\end{barticle}
\endbibitem

\bibitem[\protect\citeauthoryear{{Mondal}, {Oberoi}, and
  {Vourlidas}}{2020}]{Mondal2020a}
\begin{barticle}
\bauthor{\bsnm{{Mondal}}, \binits{S.}},
\bauthor{\bsnm{{Oberoi}}, \binits{D.}},
\bauthor{\bsnm{{Vourlidas}}, \binits{A.}}:
\byear{2020},
\batitle{{Estimation of the Physical Parameters of a CME at High Coronal
  Heights Using Low-frequency Radio Observations}}.
\bjtitle{\apj}
\bvolume{893},
\bfpage{28}.
\doiurl{https://doi.org/10.3847/1538-4357/ab7fab}.
\adsurl{2020ApJ...893...28M}.
\end{barticle}
\endbibitem

\bibitem[\protect\citeauthoryear{{Mondal} et~al.}{2019a}]{mondal_ionosphere}
\begin{bchapter}
\bauthor{\bsnm{{Mondal}}, \binits{S.}},
\bauthor{\bsnm{{Oberoi}}, \binits{D.}},
\bauthor{\bsnm{{Lonsdale}}, \binits{C.}},
\bauthor{\bsnm{{Morgan}}, \binits{J.}}:
\byear{2019}a,
\bctitle{{High precision daytime TEC measurements using the Murchison Widefield
  Array}}.
In: \beditor{\bsnm{{Doherty P.}}},
\beditor{\bsnm{{Franceschi D. G.}}} (eds.)
\bbtitle{Proceedings, 2019 URSI Asia-Pacific Radio Sci. Conference (AP-RASC
  2019)}.
\adsurl{https://www.ursi.org/proceedings/procAP19/papers2019/ionospheresurajit.pdf}.
\end{bchapter}
\endbibitem

\bibitem[\protect\citeauthoryear{{Mondal} et~al.}{2019b}]{Mondal2019}
\begin{barticle}
\bauthor{\bsnm{{Mondal}}, \binits{S.}},
\bauthor{\bsnm{{Mohan}}, \binits{A.}},
\bauthor{\bsnm{{Oberoi}}, \binits{D.}},
\bauthor{\bsnm{{Morgan}}, \binits{J.S.}},
\bauthor{\bsnm{{Benkevitch}}, \binits{L.}},
\bauthor{\bsnm{{Lonsdale}}, \binits{C.J.}},
\bauthor{\bsnm{{Crowley}}, \binits{M.}},
\bauthor{\bsnm{{Cairns}}, \binits{I.H.}}:
\byear{2019}b,
\batitle{{Unsupervised Generation of High Dynamic Range Solar Images: A Novel
  Algorithm for Self-calibration of Interferometry Data}}.
\bjtitle{\apj}
\bvolume{875},
\bfpage{97}.
\doiurl{https://doi.org/10.3847/1538-4357/ab0a01}.
\adsurl{2019ApJ...875...97M}.
\end{barticle}
\endbibitem

\bibitem[\protect\citeauthoryear{{Nindos}}{2020}]{Nindos2020}
\begin{barticle}
\bauthor{\bsnm{{Nindos}}, \binits{A.}}:
\byear{2020},
\batitle{{Incoherent Solar Radio Emission}}.
\bjtitle{Front. Astron. Space Sci.}
\bvolume{7},
\bfpage{57}.
\doiurl{https://doi.org/10.3389/fspas.2020.00057}.
\adsurl{2020FrASS...7...57N}.
\end{barticle}
\endbibitem

\bibitem[\protect\citeauthoryear{{Oberoi}, {Sharma}, and
  {Rogers}}{2017}]{oberoi2017}
\begin{barticle}
\bauthor{\bsnm{{Oberoi}}, \binits{D.}},
\bauthor{\bsnm{{Sharma}}, \binits{R.}},
\bauthor{\bsnm{{Rogers}}, \binits{A.E.E.}}:
\byear{2017},
\batitle{{Estimating Solar Flux Density at Low Radio Frequencies Using a Sky
  Brightness Model}}.
\bjtitle{\solphys}
\bvolume{292},
\bfpage{75}.
\doiurl{https://doi.org/10.1007/s11207-017-1096-1}.
\adsurl{2017SoPh..292...75O}.
\end{barticle}
\endbibitem

\bibitem[\protect\citeauthoryear{{Perley} et~al.}{2009}]{VLA2009}
\begin{barticle}
\bauthor{\bsnm{{Perley}}, \binits{R.}},
\bauthor{\bsnm{{Napier}}, \binits{P.}},
\bauthor{\bsnm{{Jackson}}, \binits{J.}},
\bauthor{\bsnm{{Butler}}, \binits{B.}},
\bauthor{\bsnm{{Carlson}}, \binits{B.}},
\bauthor{\bsnm{{Fort}}, \binits{D.}},
\bauthor{\bsnm{{Dewdney}}, \binits{P.}},
\bauthor{\bsnm{{Clark}}, \binits{B.}},
\bauthor{\bsnm{{Hayward}}, \binits{R.}},
\bauthor{\bsnm{{Durand}}, \binits{S.}},
\bauthor{\bsnm{{Revnell}}, \binits{M.}},
\bauthor{\bsnm{{McKinnon}}, \binits{M.}}:
\byear{2009},
\batitle{{The Expanded Very Large Array}}.
\bjtitle{IEEE Proc.}
\bvolume{97},
\bfpage{1448}.
\doiurl{https://doi.org/10.1109/JPROC.2009.2015470}.
\adsurl{2009IEEEP..97.1448P}.
\end{barticle}
\endbibitem

\bibitem[\protect\citeauthoryear{{Reid} and {Ratcliffe}}{2014}]{Reid2014}
\begin{barticle}
\bauthor{\bsnm{{Reid}}, \binits{H.A.S.}},
\bauthor{\bsnm{{Ratcliffe}}, \binits{H.}}:
\byear{2014},
\batitle{{A review of solar type III radio bursts}}.
\bjtitle{Res. Astron. Astrophys.}
\bvolume{14},
\bfpage{773}.
\doiurl{https://doi.org/10.1088/1674-4527/14/7/003}.
\adsurl{2014RAA....14..773R}.
\end{barticle}
\endbibitem

\bibitem[\protect\citeauthoryear{{Sokolowski} et~al.}{2017}]{Sokowlski2017}
\begin{barticle}
\bauthor{\bsnm{{Sokolowski}}, \binits{M.}},
\bauthor{\bsnm{{Colegate}}, \binits{T.}},
\bauthor{\bsnm{{Sutinjo}}, \binits{A.T.}},
\bauthor{\bsnm{{Ung}}, \binits{D.}},
\bauthor{\bsnm{{Wayth}}, \binits{R.}},
\bauthor{\bsnm{{Hurley-Walker}}, \binits{N.}},
\bauthor{\bsnm{{Lenc}}, \binits{E.}},
\bauthor{\bsnm{{Pindor}}, \binits{B.}},
\bauthor{\bsnm{{Morgan}}, \binits{J.}},
\bauthor{\bsnm{{Kaplan}}, \binits{D.L.}},
\bauthor{\bsnm{{Bell}}, \binits{M.E.}},
\bauthor{\bsnm{{Callingham}}, \binits{J.R.}},
\bauthor{\bsnm{{Dwarakanath}}, \binits{K.S.}},
\bauthor{\bsnm{{For}}, \binits{B.-Q.}},
\bauthor{\bsnm{{Gaensler}}, \binits{B.M.}},
\bauthor{\bsnm{{Hancock}}, \binits{P.J.}},
\bauthor{\bsnm{{Hindson}}, \binits{L.}},
\bauthor{\bsnm{{Johnston-Hollitt}}, \binits{M.}},
\bauthor{\bsnm{{Kapi{\'n}ska}}, \binits{A.D.}},
\bauthor{\bsnm{{McKinley}}, \binits{B.}},
\bauthor{\bsnm{{Offringa}}, \binits{A.R.}},
\bauthor{\bsnm{{Procopio}}, \binits{P.}},
\bauthor{\bsnm{{Staveley-Smith}}, \binits{L.}},
\bauthor{\bsnm{{Wu}}, \binits{C.}},
\bauthor{\bsnm{{Zheng}}, \binits{Q.}}:
\byear{2017},
\batitle{{Calibration and Stokes Imaging with Full Embedded Element Primary
  Beam Model for the Murchison Widefield Array}}.
\bjtitle{\pasa}
\bvolume{34},
\bfpage{e062}.
\doiurl{https://doi.org/10.1017/pasa.2017.54}.
\adsurl{2017PASA...34...62S}.
\end{barticle}
\endbibitem

\bibitem[\protect\citeauthoryear{{Thompson}, {Moran}, and
  {Swenson}}{2017}]{thompson2017}
\begin{bbook}
\bauthor{\bsnm{{Thompson}}, \binits{A.R.}},
\bauthor{\bsnm{{Moran}}, \binits{J.M.}},
\bauthor{\bsnm{{Swenson}}, \binits{J.} \bsuffix{George~W.}}:
\byear{2017},
\bbtitle{Interferometry and Synthesis in Radio Astronomy, 3rd Edition},
\bpublisher{{Springer Nature}},
\blocation{Cham, Switzerland}.
\doiurl{https://doi.org/10.1007/978-3-319-44431-4}.
\adsurl{2017isra.book.....T}.
\end{bbook}
\endbibitem

\bibitem[\protect\citeauthoryear{{Tingay} et~al.}{2013}]{Tingay2013}
\begin{barticle}
\bauthor{\bsnm{{Tingay}}, \binits{S.J.}},
\bauthor{\bsnm{{Goeke}}, \binits{R.}},
\bauthor{\bsnm{{Bowman}}, \binits{J.D.}},
\bauthor{\bsnm{{Emrich}}, \binits{D.}},
\bauthor{\bsnm{{Ord}}, \binits{S.M.}},
\bauthor{\bsnm{{Mitchell}}, \binits{D.A.}},
\bauthor{\bsnm{{Morales}}, \binits{M.F.}},
\bauthor{\bsnm{{Booler}}, \binits{T.}},
\bauthor{\bsnm{{Crosse}}, \binits{B.}},
\bauthor{\bsnm{{Wayth}}, \binits{R.B.}},
\bauthor{\bsnm{{Lonsdale}}, \binits{C.J.}},
\bauthor{\bsnm{{Tremblay}}, \binits{S.}},
\bauthor{\bsnm{{Pallot}}, \binits{D.}},
\bauthor{\bsnm{{Colegate}}, \binits{T.}},
\bauthor{\bsnm{{Wicenec}}, \binits{A.}},
\bauthor{\bsnm{{Kudryavtseva}}, \binits{N.}},
\bauthor{\bsnm{{Arcus}}, \binits{W.}},
\bauthor{\bsnm{{Barnes}}, \binits{D.}},
\bauthor{\bsnm{{Bernardi}}, \binits{G.}},
\bauthor{\bsnm{{Briggs}}, \binits{F.}},
\bauthor{\bsnm{{Burns}}, \binits{S.}},
\bauthor{\bsnm{{Bunton}}, \binits{J.D.}},
\bauthor{\bsnm{{Cappallo}}, \binits{R.J.}},
\bauthor{\bsnm{{Corey}}, \binits{B.E.}},
\bauthor{\bsnm{{Deshpande}}, \binits{A.}},
\bauthor{\bsnm{{Desouza}}, \binits{L.}},
\bauthor{\bsnm{{Gaensler}}, \binits{B.M.}},
\bauthor{\bsnm{{Greenhill}}, \binits{L.J.}},
\bauthor{\bsnm{{Hall}}, \binits{P.J.}},
\bauthor{\bsnm{{Hazelton}}, \binits{B.J.}},
\bauthor{\bsnm{{Herne}}, \binits{D.}},
\bauthor{\bsnm{{Hewitt}}, \binits{J.N.}},
\bauthor{\bsnm{{Johnston-Hollitt}}, \binits{M.}},
\bauthor{\bsnm{{Kaplan}}, \binits{D.L.}},
\bauthor{\bsnm{{Kasper}}, \binits{J.C.}},
\bauthor{\bsnm{{Kincaid}}, \binits{B.B.}},
\bauthor{\bsnm{{Koenig}}, \binits{R.}},
\bauthor{\bsnm{{Kratzenberg}}, \binits{E.}},
\bauthor{\bsnm{{Lynch}}, \binits{M.J.}},
\bauthor{\bsnm{{Mckinley}}, \binits{B.}},
\bauthor{\bsnm{{Mcwhirter}}, \binits{S.R.}},
\bauthor{\bsnm{{Morgan}}, \binits{E.}},
\bauthor{\bsnm{{Oberoi}}, \binits{D.}},
\bauthor{\bsnm{{Pathikulangara}}, \binits{J.}},
\bauthor{\bsnm{{Prabu}}, \binits{T.}},
\bauthor{\bsnm{{Remillard}}, \binits{R.A.}},
\bauthor{\bsnm{{Rogers}}, \binits{A.E.E.}},
\bauthor{\bsnm{{Roshi}}, \binits{A.}},
\bauthor{\bsnm{{Salah}}, \binits{J.E.}},
\bauthor{\bsnm{{Sault}}, \binits{R.J.}},
\bauthor{\bsnm{{Udaya-Shankar}}, \binits{N.}},
\bauthor{\bsnm{{Schlagenhaufer}}, \binits{F.}},
\bauthor{\bsnm{{Srivani}}, \binits{K.S.}},
\bauthor{\bsnm{{Stevens}}, \binits{J.}},
\bauthor{\bsnm{{Subrahmanyan}}, \binits{R.}},
\bauthor{\bsnm{{Waterson}}, \binits{M.}},
\bauthor{\bsnm{{Webster}}, \binits{R.L.}},
\bauthor{\bsnm{{Whitney}}, \binits{A.R.}},
\bauthor{\bsnm{{Williams}}, \binits{A.}},
\bauthor{\bsnm{{Williams}}, \binits{C.L.}},
\bauthor{\bsnm{{Wyithe}}, \binits{J.S.B.}}:
\byear{2013},
\batitle{{The Murchison Widefield Array: The Square Kilometre Array Precursor
  at Low Radio Frequencies}}.
\bjtitle{\pasa}
\bvolume{30},
\bfpage{e007}.
\doiurl{https://doi.org/10.1017/pasa.2012.007}.
\adsurl{2013PASA...30....7T}.
\end{barticle}
\endbibitem

\bibitem[\protect\citeauthoryear{{van Cappellen} et~al.}{2022}]{WSRT2021}
\begin{barticle}
\bauthor{\bsnm{{van Cappellen}}, \binits{W.A.}},
\bauthor{\bsnm{{Oosterloo}}, \binits{T.A.}},
\bauthor{\bsnm{{Verheijen}}, \binits{M.A.W.}},
\bauthor{\bsnm{{Adams}}, \binits{E.A.K.}},
\bauthor{\bsnm{{Adebahr}}, \binits{B.}},
\bauthor{\bsnm{{Braun}}, \binits{R.}},
\bauthor{\bsnm{{Hess}}, \binits{K.M.}},
\bauthor{\bsnm{{Holties}}, \binits{H.}},
\bauthor{\bsnm{{van der Hulst}}, \binits{J.M.}},
\bauthor{\bsnm{{Hut}}, \binits{B.}},
\bauthor{\bsnm{{Kooistra}}, \binits{E.}},
\bauthor{\bsnm{{van Leeuwen}}, \binits{J.}},
\bauthor{\bsnm{{Loose}}, \binits{G.M.}},
\bauthor{\bsnm{{Morganti}}, \binits{R.}},
\bauthor{\bsnm{{Moss}}, \binits{V.A.}},
\bauthor{\bsnm{{Orr{\'u}}}, \binits{E.}},
\bauthor{\bsnm{{Ruiter}}, \binits{M.}},
\bauthor{\bsnm{{Schoenmakers}}, \binits{A.P.}},
\bauthor{\bsnm{{Vermaas}}, \binits{N.J.}},
\bauthor{\bsnm{{Wijnholds}}, \binits{S.J.}},
\bauthor{\bsnm{{van Amesfoort}}, \binits{A.S.}},
\bauthor{\bsnm{{Arts}}, \binits{M.J.}},
\bauthor{\bsnm{{Attema}}, \binits{J.J.}},
\bauthor{\bsnm{{Bakker}}, \binits{L.}},
\bauthor{\bsnm{{Bassa}}, \binits{C.G.}},
\bauthor{\bsnm{{Bast}}, \binits{J.E.}},
\bauthor{\bsnm{{Benthem}}, \binits{P.}},
\bauthor{\bsnm{{Beukema}}, \binits{R.}},
\bauthor{\bsnm{{Blaauw}}, \binits{R.}},
\bauthor{\bsnm{{de Blok}}, \binits{W.J.G.}},
\bauthor{\bsnm{{Bouwhuis}}, \binits{M.}},
\bauthor{\bsnm{{van den Brink}}, \binits{R.H.}},
\bauthor{\bsnm{{Connor}}, \binits{L.}},
\bauthor{\bsnm{{Coolen}}, \binits{A.H.W.M.}},
\bauthor{\bsnm{{Damstra}}, \binits{S.}},
\bauthor{\bsnm{{van Diepen}}, \binits{G.N.J.}},
\bauthor{\bsnm{{de Goei}}, \binits{R.}},
\bauthor{\bsnm{{D{\'e}nes}}, \binits{H.}},
\bauthor{\bsnm{{Drost}}, \binits{M.}},
\bauthor{\bsnm{{Ebbendorf}}, \binits{N.}},
\bauthor{\bsnm{{Frank}}, \binits{B.S.}},
\bauthor{\bsnm{{Gardenier}}, \binits{D.W.}},
\bauthor{\bsnm{{Gerbers}}, \binits{M.}},
\bauthor{\bsnm{{Grange}}, \binits{Y.G.}},
\bauthor{\bsnm{{Grit}}, \binits{T.}},
\bauthor{\bsnm{{Gunst}}, \binits{A.W.}},
\bauthor{\bsnm{{Gupta}}, \binits{N.}},
\bauthor{\bsnm{{Ivashina}}, \binits{M.V.}},
\bauthor{\bsnm{{J{\'o}zsa}}, \binits{G.I.G.}},
\bauthor{\bsnm{{Janssen}}, \binits{G.H.}},
\bauthor{\bsnm{{Koster}}, \binits{A.}},
\bauthor{\bsnm{{Kruithof}}, \binits{G.H.}},
\bauthor{\bsnm{{Kuindersma}}, \binits{S.J.}},
\bauthor{\bsnm{{Kutkin}}, \binits{A.}},
\bauthor{\bsnm{{Lucero}}, \binits{D.M.}},
\bauthor{\bsnm{{Maan}}, \binits{Y.}},
\bauthor{\bsnm{{Maccagni}}, \binits{F.M.}},
\bauthor{\bsnm{{van der Marel}}, \binits{J.}},
\bauthor{\bsnm{{Mika}}, \binits{A.}},
\bauthor{\bsnm{{Morawietz}}, \binits{J.}},
\bauthor{\bsnm{{Mulder}}, \binits{H.}},
\bauthor{\bsnm{{Mulder}}, \binits{E.}},
\bauthor{\bsnm{{Norden}}, \binits{M.J.}},
\bauthor{\bsnm{{Offringa}}, \binits{A.R.}},
\bauthor{\bsnm{{Oostrum}}, \binits{L.C.}},
\bauthor{\bsnm{{Overeem}}, \binits{R.E.}},
\bauthor{\bsnm{{Paragi}}, \binits{Z.}},
\bauthor{\bsnm{{Pepping}}, \binits{H.J.}},
\bauthor{\bsnm{{Petroff}}, \binits{E.}},
\bauthor{\bsnm{{Pisano}}, \binits{D.J.}},
\bauthor{\bsnm{{Polatidis}}, \binits{A.G.}},
\bauthor{\bsnm{{Prasad}}, \binits{P.}},
\bauthor{\bsnm{{de Reijer}}, \binits{J.P.R.}},
\bauthor{\bsnm{{Romein}}, \binits{J.W.}},
\bauthor{\bsnm{{Schaap}}, \binits{J.}},
\bauthor{\bsnm{{Schoonderbeek}}, \binits{G.W.}},
\bauthor{\bsnm{{Schulz}}, \binits{R.}},
\bauthor{\bsnm{{van der Schuur}}, \binits{D.}},
\bauthor{\bsnm{{Sclocco}}, \binits{A.}},
\bauthor{\bsnm{{Sluman}}, \binits{J.J.}},
\bauthor{\bsnm{{Smits}}, \binits{R.}},
\bauthor{\bsnm{{Stappers}}, \binits{B.W.}},
\bauthor{\bsnm{{Straal}}, \binits{S.M.}},
\bauthor{\bsnm{{Stuurwold}}, \binits{K.J.C.}},
\bauthor{\bsnm{{Verstappen}}, \binits{J.}},
\bauthor{\bsnm{{Vohl}}, \binits{D.}},
\bauthor{\bsnm{{Wierenga}}, \binits{K.J.}},
\bauthor{\bsnm{{Woestenburg}}, \binits{E.E.M.}},
\bauthor{\bsnm{{Zanting}}, \binits{A.W.}},
\bauthor{\bsnm{{Ziemke}}, \binits{J.}}:
\byear{2022},
\batitle{{Apertif: Phased array feeds for the Westerbork Synthesis Radio
  Telescope. System overview and performance characteristics}}.
\bjtitle{\aap}
\bvolume{658},
\bfpage{A146}.
\doiurl{https://doi.org/10.1051/0004-6361/202141739}.
\adsurl{2022A&A...658A.146V}.
\end{barticle}
\endbibitem

\bibitem[\protect\citeauthoryear{{van Haarlem} et~al.}{2013}]{lofar2013}
\begin{barticle}
\bauthor{\bsnm{{van Haarlem}}, \binits{M.P.}},
\bauthor{\bsnm{{Wise}}, \binits{M.W.}},
\bauthor{\bsnm{{Gunst}}, \binits{A.W.}},
\bauthor{\bsnm{{Heald}}, \binits{G.}},
\bauthor{\bsnm{{McKean}}, \binits{J.P.}},
\bauthor{\bsnm{{Hessels}}, \binits{J.W.T.}},
\bauthor{\bsnm{{de Bruyn}}, \binits{A.G.}},
\bauthor{\bsnm{{Nijboer}}, \binits{R.}},
\bauthor{\bsnm{{Swinbank}}, \binits{J.}},
\bauthor{\bsnm{{Fallows}}, \binits{R.}},
\bauthor{\bsnm{{Brentjens}}, \binits{M.}},
\bauthor{\bsnm{{Nelles}}, \binits{A.}},
\bauthor{\bsnm{{Beck}}, \binits{R.}},
\bauthor{\bsnm{{Falcke}}, \binits{H.}},
\bauthor{\bsnm{{Fender}}, \binits{R.}},
\bauthor{\bsnm{{H{\"o}randel}}, \binits{J.}},
\bauthor{\bsnm{{Koopmans}}, \binits{L.V.E.}},
\bauthor{\bsnm{{Mann}}, \binits{G.}},
\bauthor{\bsnm{{Miley}}, \binits{G.}},
\bauthor{\bsnm{{R{\"o}ttgering}}, \binits{H.}},
\bauthor{\bsnm{{Stappers}}, \binits{B.W.}},
\bauthor{\bsnm{{Wijers}}, \binits{R.A.M.J.}},
\bauthor{\bsnm{{Zaroubi}}, \binits{S.}},
\bauthor{\bsnm{{van den Akker}}, \binits{M.}},
\bauthor{\bsnm{{Alexov}}, \binits{A.}},
\bauthor{\bsnm{{Anderson}}, \binits{J.}},
\bauthor{\bsnm{{Anderson}}, \binits{K.}},
\bauthor{\bsnm{{van Ardenne}}, \binits{A.}},
\bauthor{\bsnm{{Arts}}, \binits{M.}},
\bauthor{\bsnm{{Asgekar}}, \binits{A.}},
\bauthor{\bsnm{{Avruch}}, \binits{I.M.}},
\bauthor{\bsnm{{Batejat}}, \binits{F.}},
\bauthor{\bsnm{{B{\"a}hren}}, \binits{L.}},
\bauthor{\bsnm{{Bell}}, \binits{M.E.}},
\bauthor{\bsnm{{Bell}}, \binits{M.R.}},
\bauthor{\bsnm{{van Bemmel}}, \binits{I.}},
\bauthor{\bsnm{{Bennema}}, \binits{P.}},
\bauthor{\bsnm{{Bentum}}, \binits{M.J.}},
\bauthor{\bsnm{{Bernardi}}, \binits{G.}},
\bauthor{\bsnm{{Best}}, \binits{P.}},
\bauthor{\bsnm{{B{\^\i}rzan}}, \binits{L.}},
\bauthor{\bsnm{{Bonafede}}, \binits{A.}},
\bauthor{\bsnm{{Boonstra}}, \binits{A.-J.}},
\bauthor{\bsnm{{Braun}}, \binits{R.}},
\bauthor{\bsnm{{Bregman}}, \binits{J.}},
\bauthor{\bsnm{{Breitling}}, \binits{F.}},
\bauthor{\bsnm{{van de Brink}}, \binits{R.H.}},
\bauthor{\bsnm{{Broderick}}, \binits{J.}},
\bauthor{\bsnm{{Broekema}}, \binits{P.C.}},
\bauthor{\bsnm{{Brouw}}, \binits{W.N.}},
\bauthor{\bsnm{{Br{\"u}ggen}}, \binits{M.}},
\bauthor{\bsnm{{Butcher}}, \binits{H.R.}},
\bauthor{\bsnm{{van Cappellen}}, \binits{W.}},
\bauthor{\bsnm{{Ciardi}}, \binits{B.}},
\bauthor{\bsnm{{Coenen}}, \binits{T.}},
\bauthor{\bsnm{{Conway}}, \binits{J.}},
\bauthor{\bsnm{{Coolen}}, \binits{A.}},
\bauthor{\bsnm{{Corstanje}}, \binits{A.}},
\bauthor{\bsnm{{Damstra}}, \binits{S.}},
\bauthor{\bsnm{{Davies}}, \binits{O.}},
\bauthor{\bsnm{{Deller}}, \binits{A.T.}},
\bauthor{\bsnm{{Dettmar}}, \binits{R.-J.}},
\bauthor{\bsnm{{van Diepen}}, \binits{G.}},
\bauthor{\bsnm{{Dijkstra}}, \binits{K.}},
\bauthor{\bsnm{{Donker}}, \binits{P.}},
\bauthor{\bsnm{{Doorduin}}, \binits{A.}},
\bauthor{\bsnm{{Dromer}}, \binits{J.}},
\bauthor{\bsnm{{Drost}}, \binits{M.}},
\bauthor{\bsnm{{van Duin}}, \binits{A.}},
\bauthor{\bsnm{{Eisl{\"o}ffel}}, \binits{J.}},
\bauthor{\bsnm{{van Enst}}, \binits{J.}},
\bauthor{\bsnm{{Ferrari}}, \binits{C.}},
\bauthor{\bsnm{{Frieswijk}}, \binits{W.}},
\bauthor{\bsnm{{Gankema}}, \binits{H.}},
\bauthor{\bsnm{{Garrett}}, \binits{M.A.}},
\bauthor{\bsnm{{de Gasperin}}, \binits{F.}},
\bauthor{\bsnm{{Gerbers}}, \binits{M.}},
\bauthor{\bsnm{{de Geus}}, \binits{E.}},
\bauthor{\bsnm{{Grie{\ss}meier}}, \binits{J.-M.}},
\bauthor{\bsnm{{Grit}}, \binits{T.}},
\bauthor{\bsnm{{Gruppen}}, \binits{P.}},
\bauthor{\bsnm{{Hamaker}}, \binits{J.P.}},
\bauthor{\bsnm{{Hassall}}, \binits{T.}},
\bauthor{\bsnm{{Hoeft}}, \binits{M.}},
\bauthor{\bsnm{{Holties}}, \binits{H.A.}},
\bauthor{\bsnm{{Horneffer}}, \binits{A.}},
\bauthor{\bsnm{{van der Horst}}, \binits{A.}},
\bauthor{\bsnm{{van Houwelingen}}, \binits{A.}},
\bauthor{\bsnm{{Huijgen}}, \binits{A.}},
\bauthor{\bsnm{{Iacobelli}}, \binits{M.}},
\bauthor{\bsnm{{Intema}}, \binits{H.}},
\bauthor{\bsnm{{Jackson}}, \binits{N.}},
\bauthor{\bsnm{{Jelic}}, \binits{V.}},
\bauthor{\bsnm{{de Jong}}, \binits{A.}},
\bauthor{\bsnm{{Juette}}, \binits{E.}},
\bauthor{\bsnm{{Kant}}, \binits{D.}},
\bauthor{\bsnm{{Karastergiou}}, \binits{A.}},
\bauthor{\bsnm{{Koers}}, \binits{A.}},
\bauthor{\bsnm{{Kollen}}, \binits{H.}},
\bauthor{\bsnm{{Kondratiev}}, \binits{V.I.}},
\bauthor{\bsnm{{Kooistra}}, \binits{E.}},
\bauthor{\bsnm{{Koopman}}, \binits{Y.}},
\bauthor{\bsnm{{Koster}}, \binits{A.}},
\bauthor{\bsnm{{Kuniyoshi}}, \binits{M.}},
\bauthor{\bsnm{{Kramer}}, \binits{M.}},
\bauthor{\bsnm{{Kuper}}, \binits{G.}},
\bauthor{\bsnm{{Lambropoulos}}, \binits{P.}},
\bauthor{\bsnm{{Law}}, \binits{C.}},
\bauthor{\bsnm{{van Leeuwen}}, \binits{J.}},
\bauthor{\bsnm{{Lemaitre}}, \binits{J.}},
\bauthor{\bsnm{{Loose}}, \binits{M.}},
\bauthor{\bsnm{{Maat}}, \binits{P.}},
\bauthor{\bsnm{{Macario}}, \binits{G.}},
\bauthor{\bsnm{{Markoff}}, \binits{S.}},
\bauthor{\bsnm{{Masters}}, \binits{J.}},
\bauthor{\bsnm{{McFadden}}, \binits{R.A.}},
\bauthor{\bsnm{{McKay-Bukowski}}, \binits{D.}},
\bauthor{\bsnm{{Meijering}}, \binits{H.}},
\bauthor{\bsnm{{Meulman}}, \binits{H.}},
\bauthor{\bsnm{{Mevius}}, \binits{M.}},
\bauthor{\bsnm{{Middelberg}}, \binits{E.}},
\bauthor{\bsnm{{Millenaar}}, \binits{R.}},
\bauthor{\bsnm{{Miller-Jones}}, \binits{J.C.A.}},
\bauthor{\bsnm{{Mohan}}, \binits{R.N.}},
\bauthor{\bsnm{{Mol}}, \binits{J.D.}},
\bauthor{\bsnm{{Morawietz}}, \binits{J.}},
\bauthor{\bsnm{{Morganti}}, \binits{R.}},
\bauthor{\bsnm{{Mulcahy}}, \binits{D.D.}},
\bauthor{\bsnm{{Mulder}}, \binits{E.}},
\bauthor{\bsnm{{Munk}}, \binits{H.}},
\bauthor{\bsnm{{Nieuwenhuis}}, \binits{L.}},
\bauthor{\bsnm{{van Nieuwpoort}}, \binits{R.}},
\bauthor{\bsnm{{Noordam}}, \binits{J.E.}},
\bauthor{\bsnm{{Norden}}, \binits{M.}},
\bauthor{\bsnm{{Noutsos}}, \binits{A.}},
\bauthor{\bsnm{{Offringa}}, \binits{A.R.}},
\bauthor{\bsnm{{Olofsson}}, \binits{H.}},
\bauthor{\bsnm{{Omar}}, \binits{A.}},
\bauthor{\bsnm{{Orr{\'u}}}, \binits{E.}},
\bauthor{\bsnm{{Overeem}}, \binits{R.}},
\bauthor{\bsnm{{Paas}}, \binits{H.}},
\bauthor{\bsnm{{Pandey-Pommier}}, \binits{M.}},
\bauthor{\bsnm{{Pandey}}, \binits{V.N.}},
\bauthor{\bsnm{{Pizzo}}, \binits{R.}},
\bauthor{\bsnm{{Polatidis}}, \binits{A.}},
\bauthor{\bsnm{{Rafferty}}, \binits{D.}},
\bauthor{\bsnm{{Rawlings}}, \binits{S.}},
\bauthor{\bsnm{{Reich}}, \binits{W.}},
\bauthor{\bsnm{{de Reijer}}, \binits{J.-P.}},
\bauthor{\bsnm{{Reitsma}}, \binits{J.}},
\bauthor{\bsnm{{Renting}}, \binits{G.A.}},
\bauthor{\bsnm{{Riemers}}, \binits{P.}},
\bauthor{\bsnm{{Rol}}, \binits{E.}},
\bauthor{\bsnm{{Romein}}, \binits{J.W.}},
\bauthor{\bsnm{{Roosjen}}, \binits{J.}},
\bauthor{\bsnm{{Ruiter}}, \binits{M.}},
\bauthor{\bsnm{{Scaife}}, \binits{A.}},
\bauthor{\bsnm{{van der Schaaf}}, \binits{K.}},
\bauthor{\bsnm{{Scheers}}, \binits{B.}},
\bauthor{\bsnm{{Schellart}}, \binits{P.}},
\bauthor{\bsnm{{Schoenmakers}}, \binits{A.}},
\bauthor{\bsnm{{Schoonderbeek}}, \binits{G.}},
\bauthor{\bsnm{{Serylak}}, \binits{M.}},
\bauthor{\bsnm{{Shulevski}}, \binits{A.}},
\bauthor{\bsnm{{Sluman}}, \binits{J.}},
\bauthor{\bsnm{{Smirnov}}, \binits{O.}},
\bauthor{\bsnm{{Sobey}}, \binits{C.}},
\bauthor{\bsnm{{Spreeuw}}, \binits{H.}},
\bauthor{\bsnm{{Steinmetz}}, \binits{M.}},
\bauthor{\bsnm{{Sterks}}, \binits{C.G.M.}},
\bauthor{\bsnm{{Stiepel}}, \binits{H.-J.}},
\bauthor{\bsnm{{Stuurwold}}, \binits{K.}},
\bauthor{\bsnm{{Tagger}}, \binits{M.}},
\bauthor{\bsnm{{Tang}}, \binits{Y.}},
\bauthor{\bsnm{{Tasse}}, \binits{C.}},
\bauthor{\bsnm{{Thomas}}, \binits{I.}},
\bauthor{\bsnm{{Thoudam}}, \binits{S.}},
\bauthor{\bsnm{{Toribio}}, \binits{M.C.}},
\bauthor{\bsnm{{van der Tol}}, \binits{B.}},
\bauthor{\bsnm{{Usov}}, \binits{O.}},
\bauthor{\bsnm{{van Veelen}}, \binits{M.}},
\bauthor{\bsnm{{van der Veen}}, \binits{A.-J.}},
\bauthor{\bsnm{{ter Veen}}, \binits{S.}},
\bauthor{\bsnm{{Verbiest}}, \binits{J.P.W.}},
\bauthor{\bsnm{{Vermeulen}}, \binits{R.}},
\bauthor{\bsnm{{Vermaas}}, \binits{N.}},
\bauthor{\bsnm{{Vocks}}, \binits{C.}},
\bauthor{\bsnm{{Vogt}}, \binits{C.}},
\bauthor{\bsnm{{de Vos}}, \binits{M.}},
\bauthor{\bsnm{{van der Wal}}, \binits{E.}},
\bauthor{\bsnm{{van Weeren}}, \binits{R.}},
\bauthor{\bsnm{{Weggemans}}, \binits{H.}},
\bauthor{\bsnm{{Weltevrede}}, \binits{P.}},
\bauthor{\bsnm{{White}}, \binits{S.}},
\bauthor{\bsnm{{Wijnholds}}, \binits{S.J.}},
\bauthor{\bsnm{{Wilhelmsson}}, \binits{T.}},
\bauthor{\bsnm{{Wucknitz}}, \binits{O.}},
\bauthor{\bsnm{{Yatawatta}}, \binits{S.}},
\bauthor{\bsnm{{Zarka}}, \binits{P.}},
\bauthor{\bsnm{{Zensus}}, \binits{A.}},
\bauthor{\bsnm{{van Zwieten}}, \binits{J.}}:
\byear{2013},
\batitle{{LOFAR: The LOw-Frequency ARray}}.
\bjtitle{\aap}
\bvolume{556},
\bfpage{A2}.
\doiurl{https://doi.org/10.1051/0004-6361/201220873}.
\adsurl{2013A&A...556A...2V}.
\end{barticle}
\endbibitem

\bibitem[\protect\citeauthoryear{{Wayth} et~al.}{2018}]{Wayth2018}
\begin{barticle}
\bauthor{\bsnm{{Wayth}}, \binits{R.B.}},
\bauthor{\bsnm{{Tingay}}, \binits{S.J.}},
\bauthor{\bsnm{{Trott}}, \binits{C.M.}},
\bauthor{\bsnm{{Emrich}}, \binits{D.}},
\bauthor{\bsnm{{Johnston-Hollitt}}, \binits{M.}},
\bauthor{\bsnm{{McKinley}}, \binits{B.}},
\bauthor{\bsnm{{Gaensler}}, \binits{B.M.}},
\bauthor{\bsnm{{Beardsley}}, \binits{A.P.}},
\bauthor{\bsnm{{Booler}}, \binits{T.}},
\bauthor{\bsnm{{Crosse}}, \binits{B.}},
\bauthor{\bsnm{{Franzen}}, \binits{T.M.O.}},
\bauthor{\bsnm{{Horsley}}, \binits{L.}},
\bauthor{\bsnm{{Kaplan}}, \binits{D.L.}},
\bauthor{\bsnm{{Kenney}}, \binits{D.}},
\bauthor{\bsnm{{Morales}}, \binits{M.F.}},
\bauthor{\bsnm{{Pallot}}, \binits{D.}},
\bauthor{\bsnm{{Sleap}}, \binits{G.}},
\bauthor{\bsnm{{Steele}}, \binits{K.}},
\bauthor{\bsnm{{Walker}}, \binits{M.}},
\bauthor{\bsnm{{Williams}}, \binits{A.}},
\bauthor{\bsnm{{Wu}}, \binits{C.}},
\bauthor{\bsnm{{Cairns}}, \binits{I.H.}},
\bauthor{\bsnm{{Filipovic}}, \binits{M.D.}},
\bauthor{\bsnm{{Johnston}}, \binits{S.}},
\bauthor{\bsnm{{Murphy}}, \binits{T.}},
\bauthor{\bsnm{{Quinn}}, \binits{P.}},
\bauthor{\bsnm{{Staveley-Smith}}, \binits{L.}},
\bauthor{\bsnm{{Webster}}, \binits{R.}},
\bauthor{\bsnm{{Wyithe}}, \binits{J.S.B.}}:
\byear{2018},
\batitle{{The Phase II Murchison Widefield Array: Design overview}}.
\bjtitle{\pasa}
\bvolume{35},
\bfpage{e033}.
\doiurl{https://doi.org/10.1017/pasa.2018.37}.
\adsurl{2018PASA...35...33W}.
\end{barticle}
\endbibitem

\end{thebibliography}

\end{document}